\documentstyle[12pt,psfig]{article}

\def\EQ{\begin{equation}}
\def\EN{\end{equation}}
\def\EQA{\begin{eqnarray}}\def\ENA{\end{eqnarray}}
\def\uu{{\bf u}}
\def\kk{{\bf k}}

\begin{document}

\title{Non-locality and Intermittency in 3D Turbulence}
\author{J-P. Laval$^{1}$, B. Dubrulle$^{2,3}$,  and S. Nazarenko$^{4}$}


\maketitle

\begin{enumerate}
\item[$^1$] IGPP, UCLA, 3845 Slichter Hall, Los Angeles, CA 90095, USA
\item[$^2$] CE Saclay, F-91190 Gif sur Yvette Cedex, France
\item[$^3$] CNRS, UMR 5572, Observatoire Midi-Pyr\'en\'ees, 14 avenue Belin, F-31400 Toulouse, France
\item[$^4$] Mathematics Institute University of Warwick COVENTRY CV4 7AL, UK
\end{enumerate}

\begin{abstract}
Numerical simulations are used to determine the influence of the 
non-local and local
interactions on the intermittency corrections in the scaling properties
of 3D turbulence. We show 
that neglect of local interactions leads to an enhanced small-scale
energy spectrum and to a significantly larger
number of very intense  vortices (``tornadoes'') and stronger intermittency
(e.g. wider tails in the probability distribution functions of velocity
increments and greater anomalous corrections). On the other hand, 
neglect of the non-local interactions results in even stronger
small-scale spectrum but significantly weaker intermittency.
Thus, the amount of intermittency is not determined just by the
mean intensity of the small scales, but it is non-trivially shaped 
by the nature of the scale interactions.
Namely, the role of the non-local interactions
is to generate intense vortices responsible for intermittency and the role
of the local interactions is to dissipate them. Based on
these observations, a new  model of turbulence is proposed, in which
non-local (RDT-like) interactions couple large and small scale via a 
multiplicative process with additive noise
and the local interactions are modeled by a turbulent viscosity.  This model is
used to derive a simple toy version of the Langevin equations for 
small-scale velocity
increments. A Gaussian approximation for the large scale fields 
yields the Fokker-Planck equation for the probability distribution 
function of the velocity increments. Steady state solutions of this 
equation allows to qualitatively
explain the anomalous corrections and the skewness generation along 
scale. A crucial role is played by the correlation between the 
additive and the
multiplicative (large-scale) process, featuring the correlation between
the stretching and the vorticity.

\end{abstract}

\newpage


\vspace{0.1cm}

A puzzling feature of three-dimensional turbulence are the large deviations
from Gaussianity observed as one probes smaller and smaller scales. These
deviations are usually  believed to be associated with the spatial
intermittency
of small-scale structures, organized into very thin and elongated intense
vortices (``tornadoes'')\cite{siggia81,vincent91,she90}. They are 
responsible for anomalous
corrections
to the normal scaling behavior of structure functions associated
with the Kolmogorov 1941 \cite{kolmogorov41} picture of turbulence. 
In this picture,
energy containing structures (the so-called "eddies") at a given scale,
interact with other eddies of smaller but comparable size to transfer
energy at a constant rate down to the dissipative scale. A simple prediction
of this local picture of turbulence is the famous $k^{-5/3}$ energy spectrum,
which has been observed in numerous high Reynolds number experimental data
  and numerical simulations. The local theory of turbulence further
leads to the prediction that the $n^{th}$ moment of a velocity
increment $\delta u_\ell =u(x+\ell)-u(x)$ over a distance $\ell$ should scale
like $\ell^{n/3}$. This behavior has never been observed in turbulent flows,
and is it now widely admitted that anomalous corrections exist at any finite
Reynolds number. Various scenarii have been so far proposed to explain and
compute the anomalous corrections. To mention but a few: spatial
intermittency of the energy dissipation \cite{kolmogorov62}, multi-fractal
scaling \cite{parisi85}, large deviations of multiplicative cascades
\cite{frischbook95}, extremum principle \cite{castaing89},
zero-modes of differential operators \cite{gawedski95},
scale covariance \cite{dubrulle98b} .
These approaches all try to model the breakdown of the exact local scale
invariance underlying the Kolmogorov 1941 picture. In a recent study
of finite size effects, Dubrulle \cite{dubrulle00} showed that some
properties of the structure functions (non-power law behavior, nonlinear
exponents,..) could be explained within a framework in which
finite size cut-offs play a central role, and are felt throughout
the so-called "inertial range". Such finding is in clear contradiction
with the "local" K41 theory, in which eddies in the inertial range
are insensitive to the UV and IR end of the energy spectrum, and
only interact with their neighbors (in the scale space) via
"local" interactions (involving triads of comparable size). This
observation motivated us to consider a new scenario for turbulence, in
which anomalous corrections and deviations from Gaussianity
are the result of {\sl non-local interactions} between energy containing
structures. By non-local interactions, we mean interactions between
well separated scales (or highly elongated wavenumber triads). 
A numerical analysis on the role of the different triadic interactions 
in the energy cascade have been previously done 
by Brasseur et al \cite{brasseur94} and Domaradzkiet al \cite{domaradzki94}.
We have recently demonstrated via high resolution numerical simulations
that in 2D turbulence, the small scale dynamics is essentially
governed by their non-local interactions
with the large  scales \cite{laval99}, \cite{laval98}.
This feature seems natural in view of the large-scale condensation
of vortices \cite{borue93}. As a result, the weak small scales are more
influenced by the strong large-scale advection and shearing, than by
mutual interactions between themselves. This makes non-local
interactions the dominant process at small scales.\

2D turbulence is very special, because there is no vortex stretching.
As a result, there is no increase of vorticity towards smaller scales
as is observed in 3D turbulence. A quantitative feature underlying this
difference is the shape of the energy spectra at small scales: it is
$k^{-3}$ in 2D turbulence, much steeper than $k^{-5/3}$ 3D energy
spectra. In fact,  a simple estimate shows
that the borderline between local and non-local
behaviors is precisely this $k^{-3}$ spectrum: only for energy spectra
steeper than $k^{-3}$  can one prove rigorously than the dominant interactions
are non-local. It is therefore not our intention to claim that 3D turbulence
is non-local, and in fact we do believe that local interactions
are responsible for the $k^{-5/3}$ energy spectra. However, it is not 
unreasonable
to think  that evolution of the higher cumulants (responsible
for deviations from Gaussianity at large deviations) is more
non-local than it is for the energy spectrum.
Indeed, calculation of higher cumulants in Fourier space involves
integrations over larger sets of wavenumbers  which
(even if close in values pairwise) cover larger range of
scales than in the case of lower order cumulants.\

A natural tool to study the role of the non-local interactions
is the numerical simulation, because it allows
a direct check of their influence by artificial
switching-off of the elongated wavenumber triads
in the Navier-Stokes equations or, on the contrary, retaining
only such triads.  A limitation of
this approach lies in the restricted range of Reynolds number we are
able to simulate. However,  a recent comparison
\cite{arneodo96} showed that anomalous
corrections and intermittency effects are quite insensitive to the Reynolds
number. This shows relevance of a
low Reynolds number numerical study of intermittency.
This approach is detailed in the next section,
where we
examine the dynamical role of the local interactions
at the small scales using a simulation
in which these interactions have been removed. We show that, as compared
 with a full simulation of the Navier-Stokes
equations, such ``nonlocal'' simulation exhibits a flatter spectrum
at small scales and stronger intermittency. As a qualitative indicator
of intermittency we use plots of the vortex structures whose intensity
greatly exceeds the r.m.s. vorticity value (``tornadoes'') and PDF's plots, 
whereas to quantify intermittency we measure the structure function
scaling exponents. We show that  main effect of the local interactions 
can be approximately described by a
turbulent viscosity, while the non-local interactions are responsible for
development of the localized intense vortices and the deviations of 
Gaussianity. To validate that the enhanced intermittency is not
merely a result of the  increased  mean  small-scale intensity
(which could be also caused by some reason other than the non-locality)
we perform another numerical experiment in with the elongated triads were
removed. Such ``local'' experiment resulted in even stronger small-scale
intensity but the intermittency became clearly weaker.
In the Section 2, we show how our findings can be
used to provide both a qualitative estimate of the intermittency
exponents, and
a derivation of the Langevin equations for the velocity
increments.\

\section{Non-local interactions in 3D turbulence}
\subsection{The problematics}

We consider the Navier-Stokes equations:
\EQ
\partial_t {\bf u}+{\bf u}\cdot {\bf \nabla}
{\bf u}=-{\bf \nabla} p+\nu \Delta {\bf u}+{\bf f},
\label{navierstart}
\EN
where ${\bf u}$ is the velocity, $p$ is the pressure, $\nu$ is the molecular
viscosity and ${\bf f}$ is the forcing. In a typical situation,
the forcing is provided by some boundary conditions (experiments)
or externally fixed, e.g. by keeping  a fixed low-wavenumber Fourier
mode at a constant amplitude (numerical simulation).
This situation typically gives rise to
quasi-Gaussian large-scale velocity fields, while  smaller-scale
velocities display increasingly non-Gaussian statistics.
  Presence of the forcing guarantees  existence
of a stationary steady state in which the total energy is constant.
In absence of forcing, the turbulence energy decays steadily, due to
losses through viscous effects. However, starting from a 
quasi-Gaussian large-scale field,
one can still observe  development of
increasingly non-Gaussian small scales in the early stage of the
decay.
  We will study the effect of the non-local interactions on the 
statistics of such
non-Gaussian small scales. For this,
we introduce a filter function $G({\bf x})$ in order to separate
the large and small scales of the flow.
In our numerical procedure, the filter $G$ will be taken as
a cut-off. We have checked that the results are insensitive to
the choice of the filter, provided the latter decays fast enough at infinity.
Using the filter, we
decompose the velocity field into large scale and
small scale components:
\EQA
{\bf u}({\bf x},t)&=&{\bf U}({\bf x},t)+{\bf u'}({\bf x},t),\nonumber\\
{\bf U}({\bf x},t)&\equiv&\overline{u}
=\int G({\bf x}-{\bf x'}){\bf u}({\bf x'},t)d{\bf x'}.
\label{decompo}
\ENA
Equations for the large scales of motion are obtained by
application of the spatial filter (\ref{decompo}) to the individual terms
of the basic equations (\ref{navierstart}). They are:
\EQA
\partial_j U_j&=&0,\nonumber\\
\partial_t U_i+\partial_j \overline{U_i U_j}
+\partial_j\overline{U_i u_j+U_j u_i}+\partial_j
\overline{u_i u_j}&=& \nonumber \\
-\partial_i P
+\nu\Delta U_i+F_i,
\label{LSEqint}
\ENA
In these equations, we have dropped primes on sub-filter components for 
simplicity;
this means that from now on, any large-scale quantities are denoted by
a capital letter, while the small-scale quantities are denoted by a 
lower case letter.
Equation for the small-scale component is obtained by  subtracting 
the large-scale
equation from the basic equations (\ref{navierstart}); this gives
\EQA
\partial_j u_j=0,\nonumber\\
\partial_t u_i+\partial_j\left(
(U_i +u_i)(U_j+u_j)-\overline{(U_i+u_i)(U_j+u_j)}\right)
&=& \nonumber\\
-\partial_i p
+\nu\Delta u_i+f_i,
\label{ssEqint}
\ENA
Several terms contribute to the  interaction of scales: {\sl non-local}
terms, involving the product of a large scale and a small scale
component, and a {\sl local} term, involving two small-scale components.
One way to study  the dynamical effect of these contributions at small
scale,
is to  integrate numerically a set of
two coupled equations, in which the local small scale interactions have been
switched off at the small scales \footnote{We do not switch off
the local interactions at  the large scales: this would hinder the 
cascade mechanism and
prevent small scale generation from  an initial large scale field.}.
This corresponds to the following
set of equations,
\EQA
\partial_t U_i + \partial_j \overline{U_i U_j}
  &=& - \partial_j\overline{U_i u_j} - \partial_j\overline{U_j u_i}
      - \partial_j\overline{u_i u_j} \nonumber \\
  &&-\partial_i P +\nu\Delta U_i+f_i\\
\partial_t u_i +\partial_j (U_i u_j)+\partial_j (u_i U_j)
&=& -\partial_i p +\nu\Delta u_i+\sigma_i, \label{eq:seteqs} \\
\partial_j U_j=\partial_j u_j &=&  0 \nonumber,
\ENA
where
\EQ
\sigma_i=\partial_j \left(\overline{U_i U_j}-U_i U_j+\overline{u_j 
U_i}+\overline{U_j u_i}\right).
\label{pseudoforce}
\EN
The later describes a forcing of the small scales by the large scales via the
energy cascade mechanism. This term is always finite even when the external
forcing $\bf{f}$ (which is always at large scales) is absent.
The small scale equation is linear and it resembles the equations of the
Rapid Distortion Theory \cite{townsend76}. We shall therefore refer
to this new model as the RDT model.
The corresponding solution was then compared with a reference simulation
performed at the same resolution, with the same initial condition.
Note that this comparison is rather expensive numerically: splitting
the equations of motions between resolved and sub-filter component
leads to additional Fourier transforms, and increases the
computational time by a factor 3. This sets a practical limitation
to the tests we could perform on our workstation. Also,
an additional limitation came from the need of scale
separation between the "large" and the "small" scales. This scale
separation is mandatory in order to define "non-local" interactions.
Their influence on anomalous scaling can be checked only
if the typical small scale lies within the inertial range. For this
reason, we were led to consider a situation of decaying (unforced)
3D turbulence, with a flat energy spectrum at large scale, and
an "inertial range" mostly concentrated at small to medium scales
($10<k<40$ for a $256^3$ simulation). Indeed, at the resolution we could
achieve, forced turbulence developed an inertial range of scale
around $k=8$, too small for the scale separation to be effective
(see \cite{laval00a} for a study and discussion of this case and its relevance
to LES simulations).
Decaying turbulence does not, by definition, achieve a statistically stationary
state, with mathematically well defined stationary probability distribution 
functions (PDF). Therefore, all the PDF were computed at a fixed time which we have chosen to be at the end of each simulation (at t=0.48).

\subsection{The numerical procedure}

\subsubsection{The numerical code}
Both the Navier Stokes equation (\ref{navierstart}) and the set of RDT
  equations (\ref{eq:seteqs}) were integrated with a pseudo-spectral code
  (see \cite{vincent91} for more details on the code). In the RDT
  case, a sharp cut-off in Fourier space was used to split the velocity
  field into large and small-scale components in Fourier space and all
  non-linear terms were computed separately in the physical space.
  The aliasing was removed by keeping only the $2/3$ largest modes
  corresponding to the 85 first modes in our case. The calculations
  presented here were done with $256^3$ Fourier modes and a viscosity
  of $1.5\,10^{-3}$ corresponding to a Reynolds number 
$57<R_{\lambda}<80$ (where
  $R_{\lambda}$ is the Reynolds number based on the Taylor micro-scale 
$\lambda$).

\subsubsection{The simulations}
The test was performed in a situation of decaying turbulence
  were the forcing term $f_i$ is set to zero.
  The initial condition was computed by a direct simulation
  using a Gaussian velocity
  distribution as a starting field. We have chosen the velocity field after
  several turnover times to be the initial condition for our simulations:
  the direct numerical simulation (DNS), the RDT simulation
  (eq. \ref{eq:seteqs}) and the ``local'' experiment (see the end of this
section). In order to allow  enough energy
at the large scales, the sharp cut-off
filter was taken at the wavenumber $k=24$ corresponding to approximatly 
5 Kolmogorov scales. Because of the very low speed of the RDT simulation
  (three times more expensive than
   the DNS) the two simulations were performed
between t=0 and t=0.48 corresponding to approximately 2.5 turnover times.

\subsection{Comparison of the RDT and DNS experiments}
\subsubsection{Spectra}

\begin{figure}[hhh]
\centerline{\psfig{file=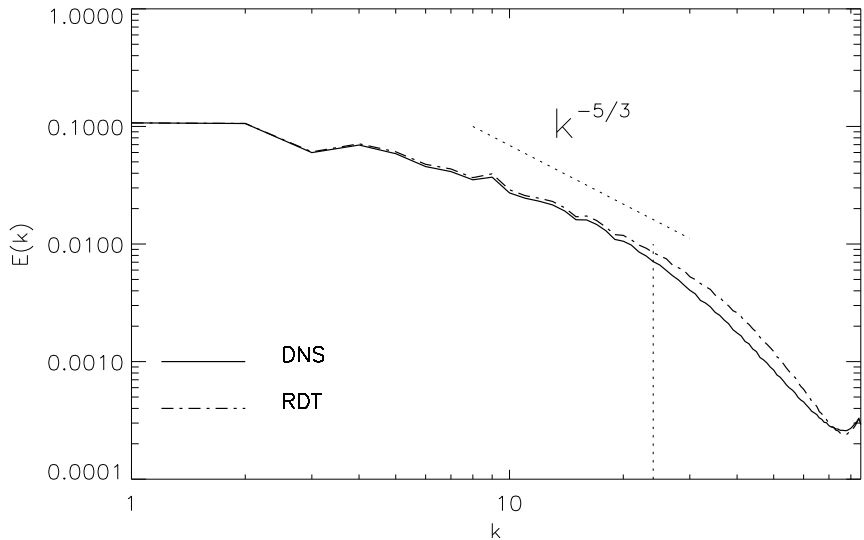,width=12cm}}
\caption[]{Comparison of energy spectra at t=0.48 obtained via the DNS and 
the RDT simulation.}
\label{fig:compengspctdnsrdt}
\end{figure}

The comparison of spectra is shown in Fig \ref{fig:compengspctdnsrdt}.
In the DNS case, one observes a classical evolution, in which
the large scale energy decays while the inertial $k^{-5/3}$ range 
tends to move towards smaller scales. It can be seen  that the inertial
 range (characterized by the $-5/3$ slope)
only marginally exists.
In the RDT case, one observes a similar evolution at large scale, while a
tendency towards a flatter spectrum is observed near and beyond the
separating scale (beyond which local interactions are ignored). We checked that
this behavior is not sensitive to the resolution.

The range of computed scales
is insufficient to find a reliable value of the spectral slope in the RDT case.
However, this slope can be predicted by a simple dimensional argument.
This argument was presented for the 2D case in
\cite{nazarenko98}, but it is essentially
the same for 3D.
Indeed, in the RDT case the small-scale equations are linear and, 
therefore, the energy
spectrum $E(k)$ must be linearly proportional to the energy dissipation rate,
$\epsilon$. In this case,  the only extra dimensional parameter
(in comparison with the local/Kolmogorov case) is  the large-scale rate of
strain $\alpha$. There is the only dimensional combination of 
$\epsilon$, $\alpha$
and wavenumber $k$ that has the dimension of $E(k)$; this gives
\begin{equation}
E_k = {C \epsilon \over \alpha} k^{-1},
\end{equation}
where $C$ is a non-dimensional constant.
In our case, our resolution is too low
  to be able to check whether the RDT spectrum follows a
$k^{-1}$ law, but we clearly see the tendency to a flatter than $-5/3$ slope.
The RDT case is reminiscent of the boundary layer, in
which a $k^{-1}$ spectra have been observed \cite{perry86}.
This is not surprising because presence of the mean
shear increases the non-locality of the scale interactions 
corresponding to  RDT.
In fact, an exact RDT analysis of the shear flow does predict formation
of the   $k^{-1}$ spectrum \cite{nazarenko00}.

\subsubsection{Structures}
3D turbulence is characterized by intense thin vortex
filaments (``tornadoes'') \cite{siggia81,vincent91,vincent94,she90}.
Their radii are of order of the dissipative (Kolmogorov) scale which
in this case is determined by the balance of the large-scale straining
and viscous spreading. In this respect, these vortices are similar to
the classical Burgers vortex solution.
\begin{figure}[hhh]
\centerline{\psfig{file=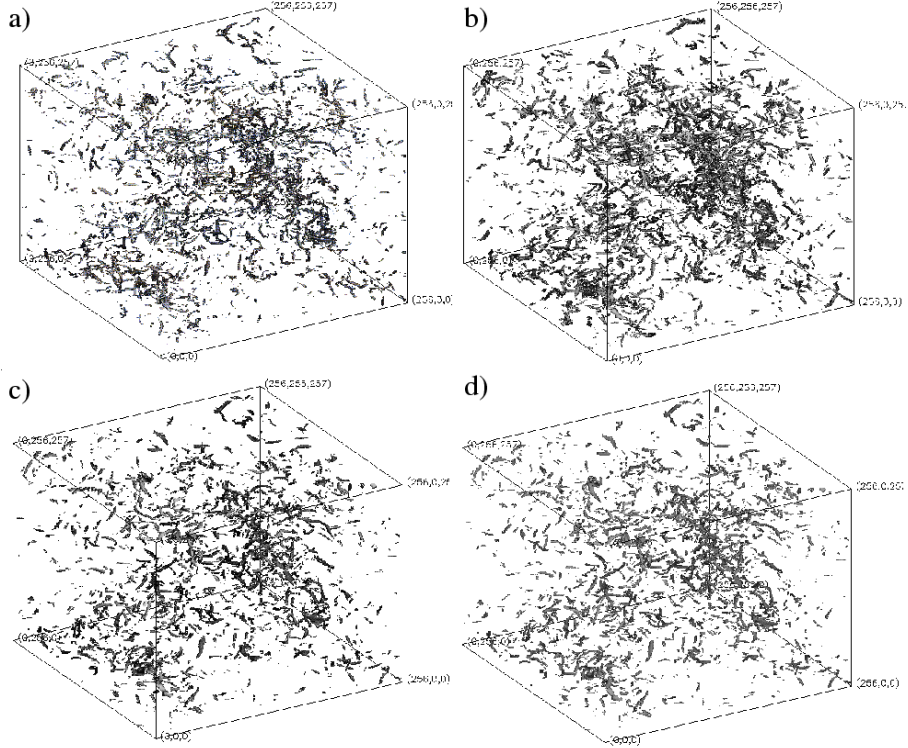,angle=-90,width=12cm}}
\caption[]{Comparison of vortex structures (isoplot of vorticity
  $\vert \omega \vert = 3.5 \; \omega_{RMS} = \omega_c $ at t=0.48) for a) DNS, b) RDT,
  c) RDT + constant turbulent viscosity, d) RDT + the RNG-type turbulent 
viscosity. The number of ``tornedoes'' estimated by $N=\int_{\omega_c}^{\infty} pdf(\vert \omega\vert) \; d\omega$ are respectively 8245 (a), 21669 (b), 10226 (c) and 10779 (d).}
\label{fig:compstructdnsrdt}
\end{figure}
  Fig. \ref{fig:compstructdnsrdt} shows a comparison
of the ``tornadoes''  observed in the DNS and in the RDT which visualized
by plotting surfaces of strong vorticity ($\vert\omega\vert > 3.5 \; \omega_{rms}$). In both cases, thin filamentary structures are observed, but they
appear to be much more numerous in the RDT case. Obviously, local interactions
tend to dissipate the ``tornadoes'' which can be interpreted as a
  mutual distortion and entanglement of ``tornadoes''
preventing their further stretching by the large scales.
On the macroscopic level, this can be regarded as an additional,
``turbulent'', viscosity produced by the local interactions. This is
compatible with the flattening of the energy spectrum in the RDT case, which
we interpreted above in terms of the turbulent viscosity effect.

It is interesting that the  Burgers vortex is essentially
a linear solution because of the cylindrical shape of this vortex which
prevents appearance of the quadratic (in vorticity) terms. Such a  linearity 
is a typical feature of all RDT solutions. On the other hand, there is 
another candidate which has often been considered to be responsible for intermittency:
this is a vortex reconnection process which is believed to lead to a
finite time singularity formation (at least for inviscid fluids). 
Note that the vortex reconnection is an essentially nonlinear process in
which the local scale interactions are playing an important role and
cannot be ignored. Indeed, there is no finite time blow-up solutions in linear RDT.
Likewise, the vorticity grows only exponentially in the  Burgers vortex and it
does not blow up in a finite time. From this perspective, our numerical results show 
that the local (vortex-vortex) interactions mostly lead to destruction of 
the intense vortices and prevention of their further Burgers-like (exponential)
growth which has a negative effect on the intermittency. This process seems to overpower
the positive effect of the local interactions on the intermittency which is 
related to the reconnection blow-ups. At the moment, it is not possible to say
if the same is true for much higher Reynolds number flows.

\subsubsection{Turbulent viscosity}

\begin{figure}[hhh]
\centerline{\psfig{file=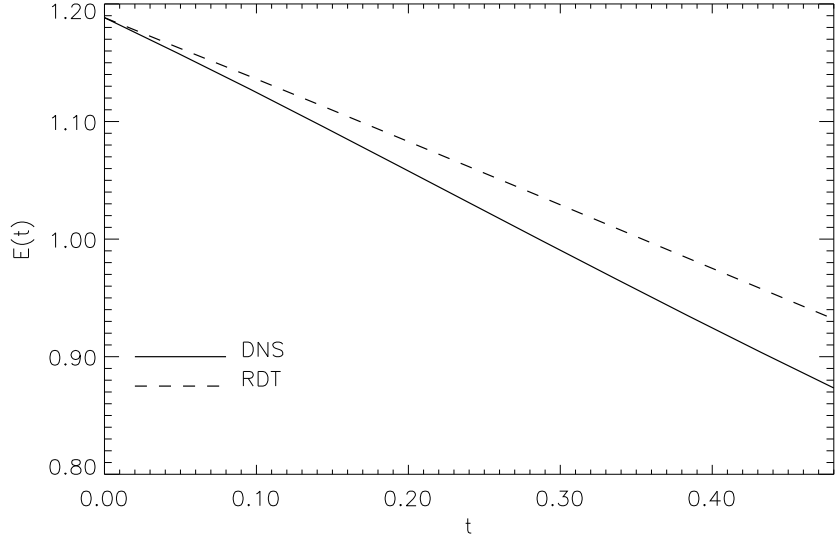,width=12cm}}
\caption[]{Comparison of the time evolution of total energy.}
\label{fig:compengdnsrdt}
\end{figure}

\begin{figure}[hhh]
\centerline{\psfig{file=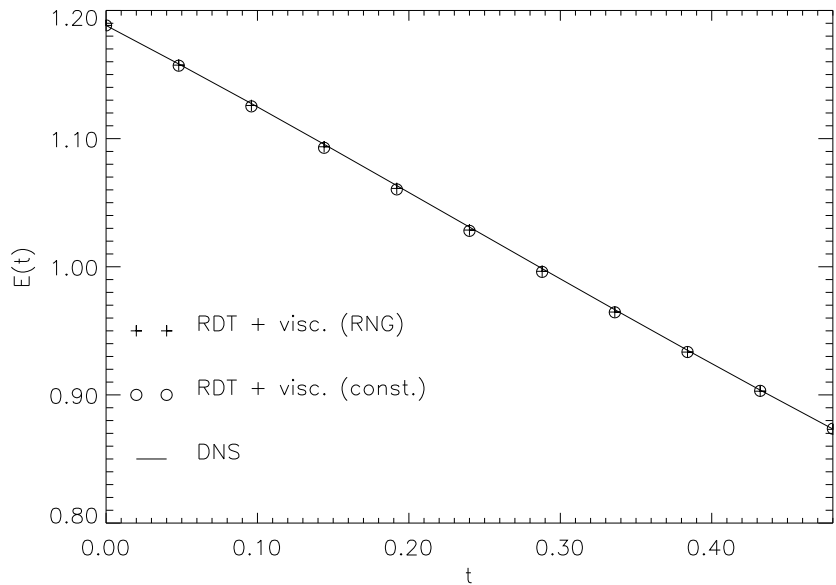,width=12cm}}
\caption[]{Time evolution of total energy obtained via the DNS
and the RDT simulation with the two different turbulent viscosities.}
\label{fig:compengdnsrdtv}
\end{figure}

Comparison of the  DNS and RDT results for the time evolution of
the total energy is shown in 
  (Fig. \ref{fig:compengdnsrdt}). One clearly observes a slower
decline of the total energy in the RDT case, as if there was a lower
viscosity.
 This results is not surprising: it is well known
  that the influence of energy motions
onto well separated large scales (see \cite{kraichnan76,dubrulle91b,yakhot86}
for systematic expansions) is through an effective eddy viscosity,
supplied by the $<uu>$ term.
Our result suggests that, to a first approximation, the difference
between the RDT and the DNS could be removed by including an additional
"turbulent" viscosity in the RDT simulation.
\begin{figure}[hhh]
\centerline{\psfig{file=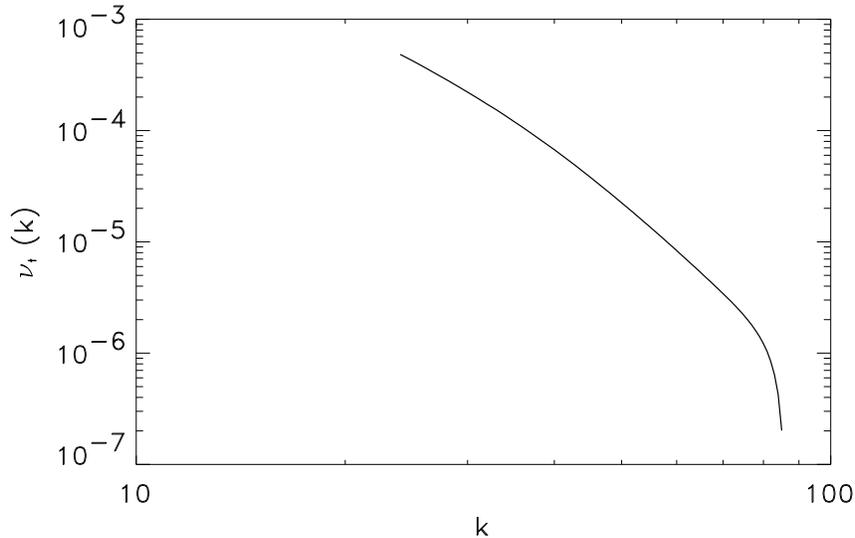,width=12cm}}
\caption[]{Spectrum of the turbulent viscosity
  computed by (\ref{viscocanuto}) at t=0.48}
\label{fig:spectnut}
\end{figure}
 For the sake of simplicity,
we decided to choose an isotropic tensor, chosen as to conserve the total
energy. We tried two simple viscosity prescription: one in which $\nu_t$ is
constant, and one (Fig. \ref{fig:spectnut}) in which the viscosity prescription follows the
shape dictated by Renormalization Group Theory  (see e.g. \cite{canuto96} ):

\begin{equation}
\nu_t(k) = \left( \nu^2 + A \int_{k}^{+\infty} q^{-2} E(q) dq \right)^{1/2} -\nu
\label{viscocanuto}
\end{equation}
The constants were adjusted as to obtain a correct
energy decay (Fig. \ref{fig:compengdnsrdtv}).
They are $\nu_t=0.0002$ for the constant $\nu_t$ prescription, and $A=0.02$ 
for the other one.
We elaborate more on the choice of this turbulent viscosity in Section
\ref{sec:turbvisc}.  Yet another method we tried was to replace
the neglected nonlinear term (interaction of small scales among themselves)
with its mean value. Dividing this mean nonlinear term by $k^2$ for each 
$k$ one can compute the turbulent viscosity $\nu_t(k)$. 
The result is interesting: $\nu_t$ turns out to be nearly independent of 
$k$; this provides an extra justification for the simple model in which
 $\nu_t=$ const.
 
\begin{figure}[hhh]
\centerline{\psfig{file=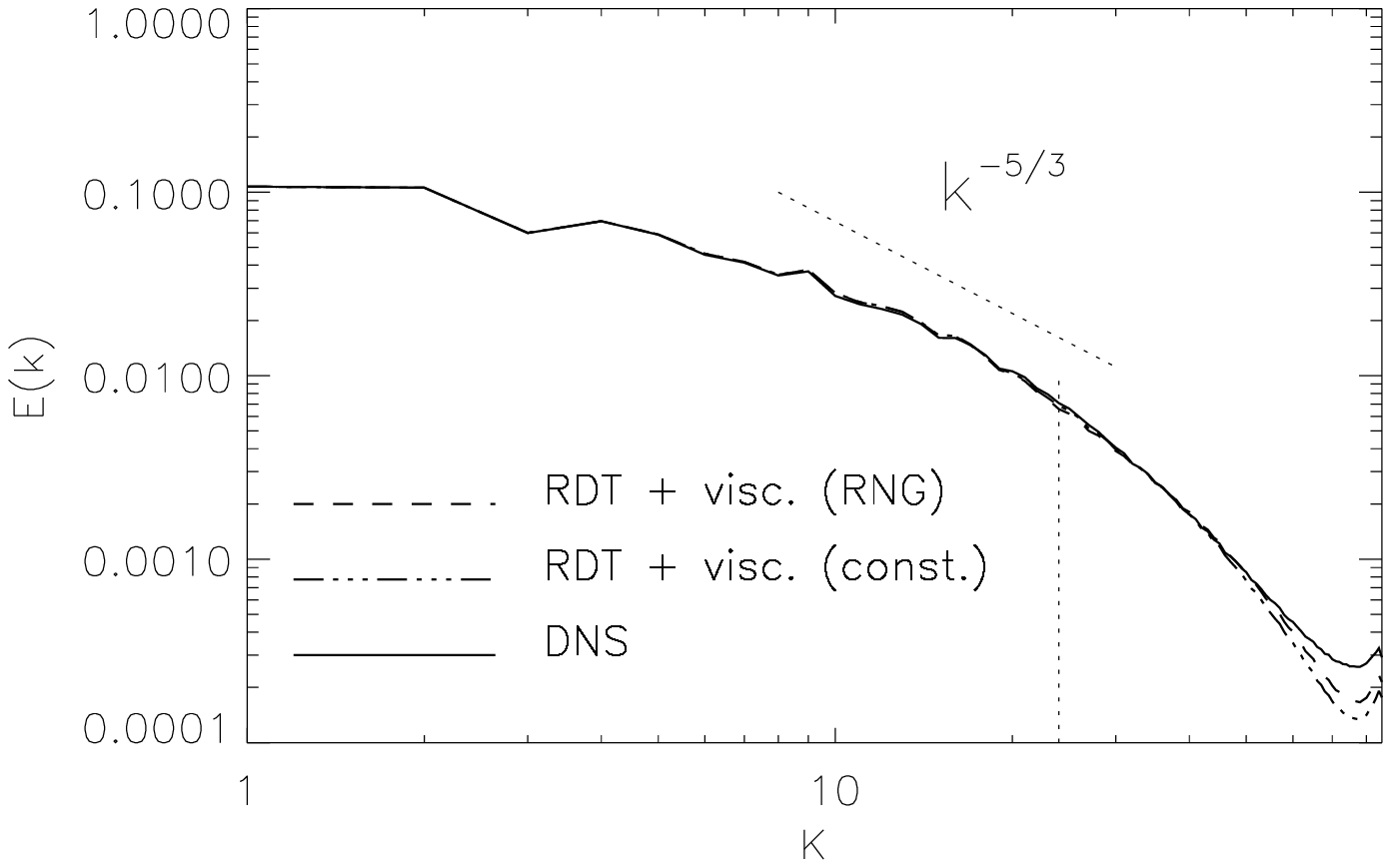,width=12cm}}
\caption[]{Comparison of energy spectra obtained via the DNS and the RDT 
simulation with two different forms of the
turbulent viscosity}
\label{fig:compengspectdnsrdtv}
\end{figure}

The energy spectra and energy decay
obtained with this new RDT simulation are shown on Figs.
  \ref{fig:compengspectdnsrdtv} and \ref{fig:compengdnsrdtv}.
One sees that one now captures exactly the energy decay of the DNS.
The energy spectra  become closer to the DNS result at low and
intermediate $k$, whereas at the
high $k$ they depart from DNS indicating an over-dissipation  of the
smallest scales. The later is an artifact of our crude choice for the turbulent
viscosity ignoring its anisotropy and possibility for it to take 
negative values.
The turbulent viscosity also influences the anomalous corrections,
as we will now show it.

\subsubsection{PDF's and exponents}

\begin{figure}[hhh]
\centerline{\psfig{file=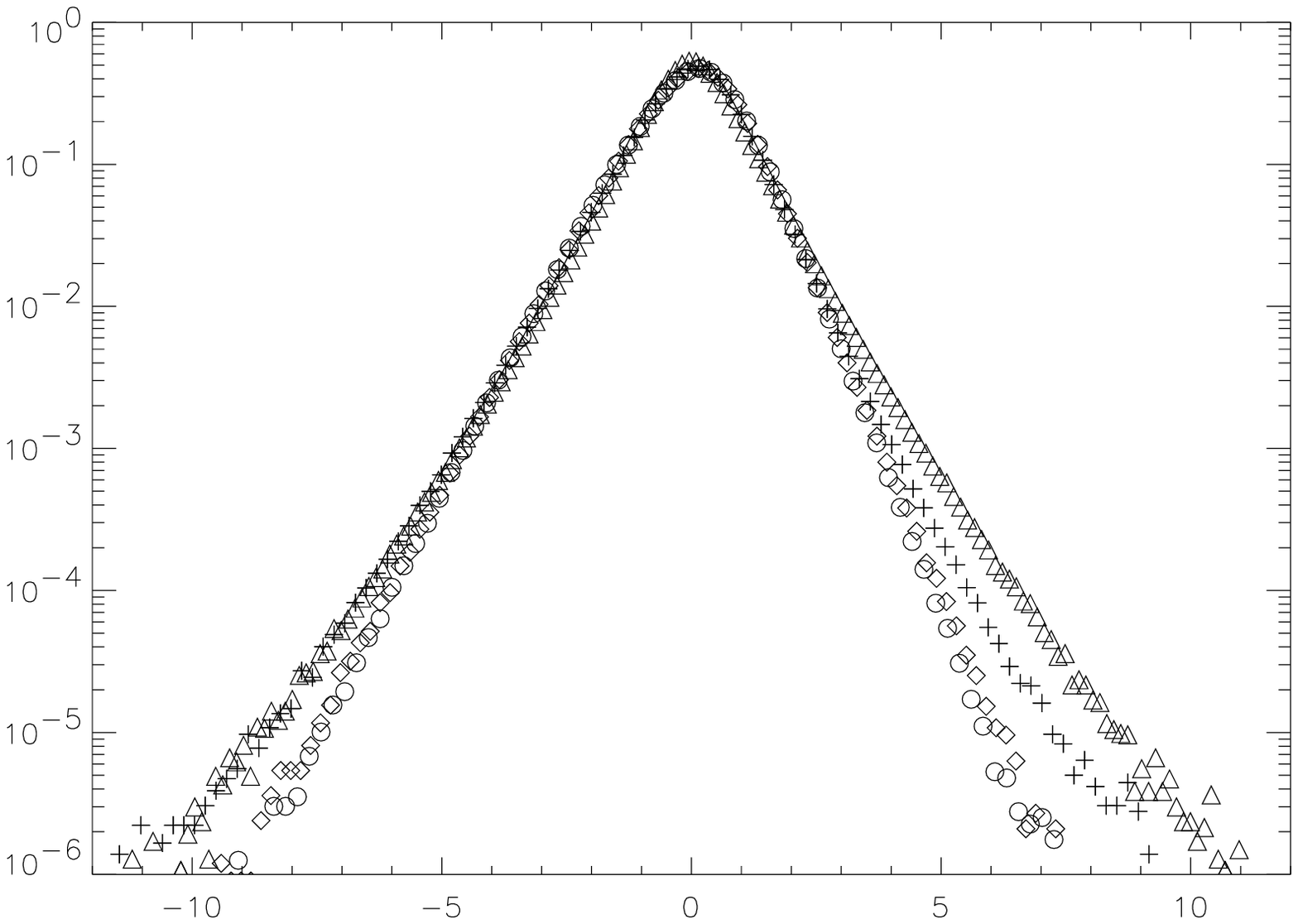,width=12cm}}
\caption[]{Comparison of the PDF of the longitudinal velocity 
increments defined by Eq. (\ref{increments}) for  $\ell=2\pi/256$ 
($5. \, 10^7$ statistics  at $t=0.48$ for the velocity field from the four 
simulations: DNS (circle), RDT (crosses), RDT + constant viscosity (diamonds)
 and RDT + 
viscosity computed from RNG (triangles)   }
\label{fig:pdfl1}
\end{figure}

\begin{figure}[hhh]
\centerline{\psfig{file=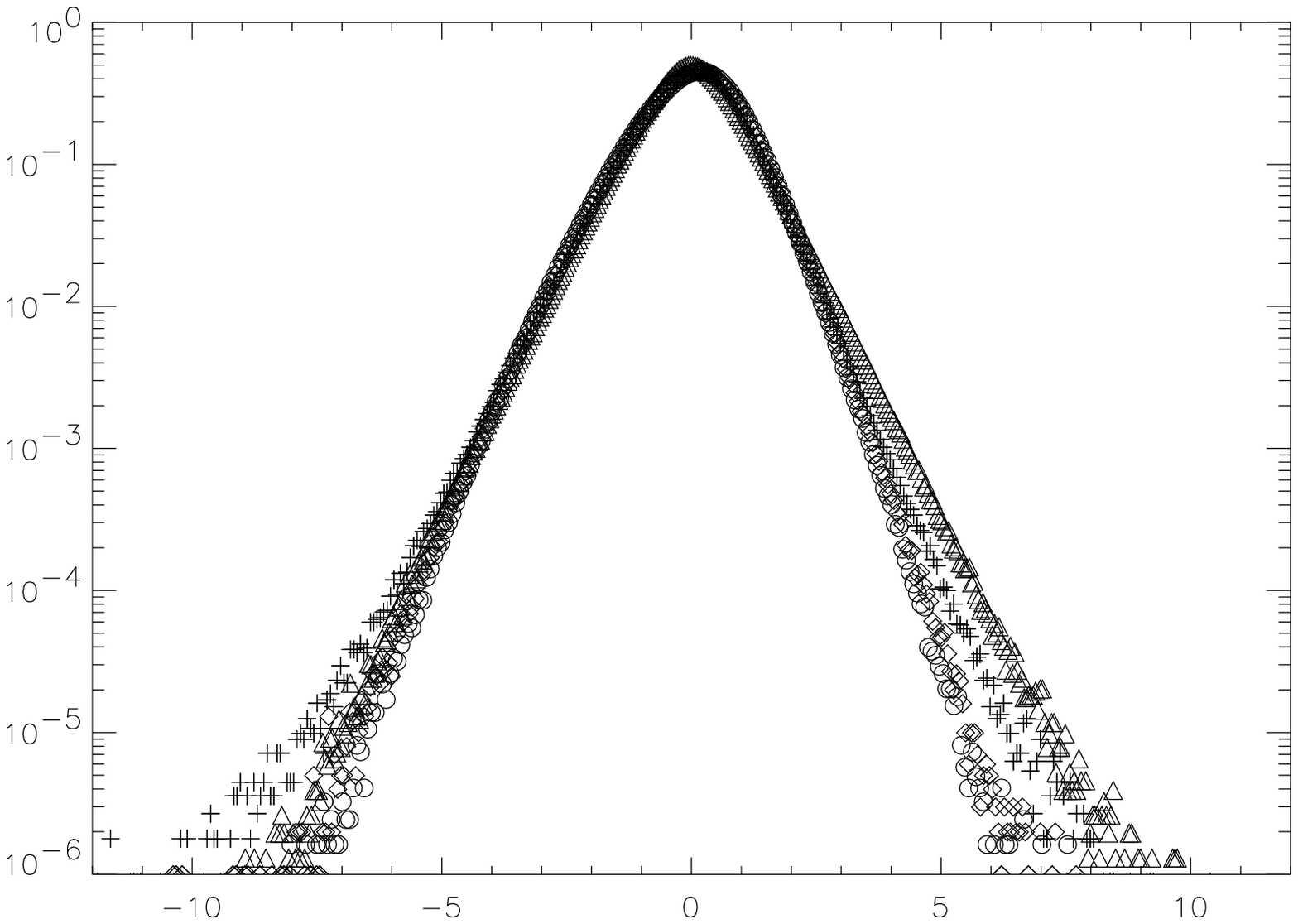,width=12cm}}
\caption[]{Same caption as in Fig. \ref{fig:pdfl1} for $\ell=2\pi/64$}
\label{fig:pdfl4}
\end{figure}

\begin{figure}[hhh]
\centerline{\psfig{file=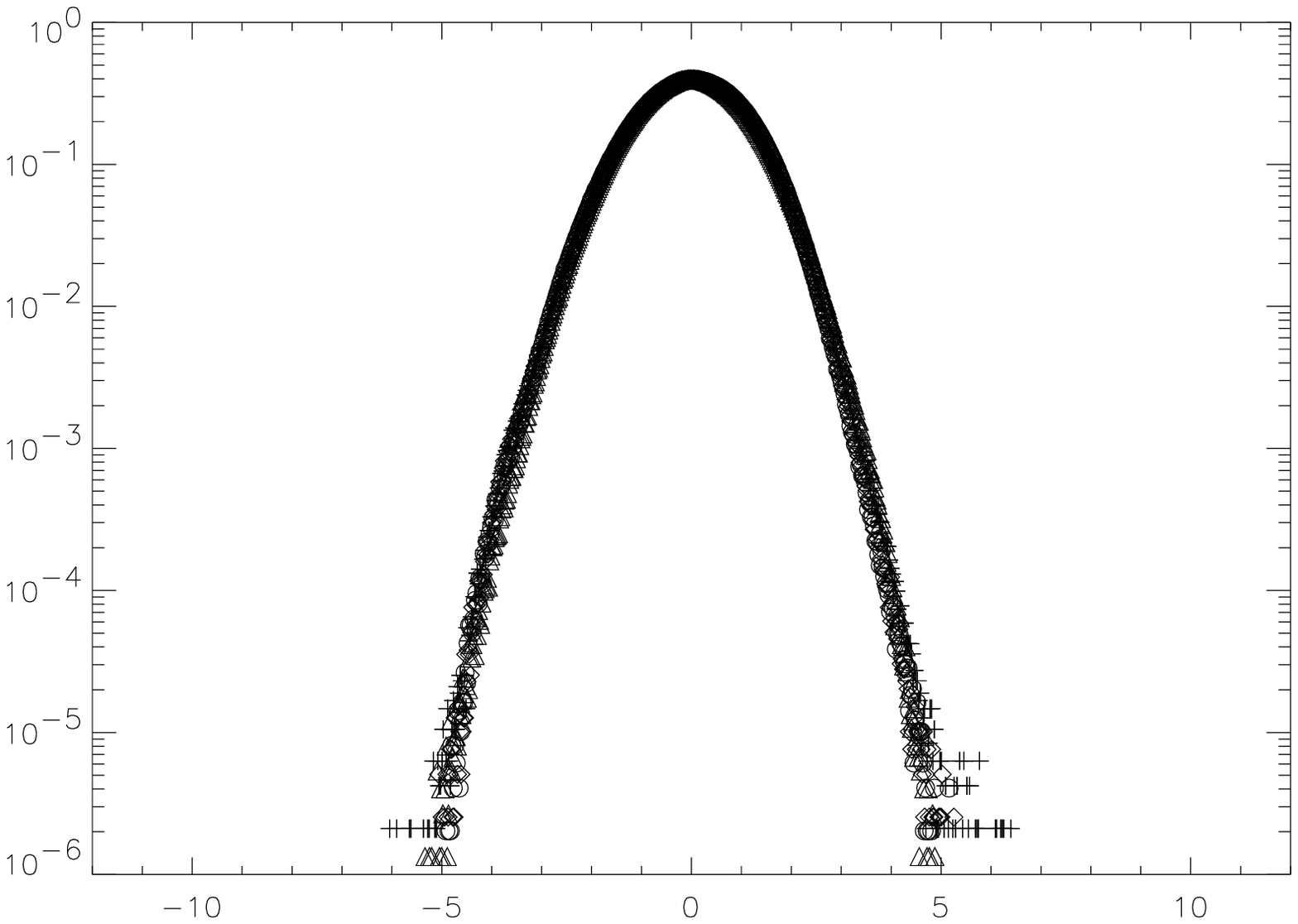,width=12cm}}
\caption[]{Same caption as in Fig. \ref{fig:pdfl1} for $\ell=2\pi/4$}
\label{fig:pdfl12}
\end{figure}

\begin{figure}[hhh]
\centerline{\psfig{file=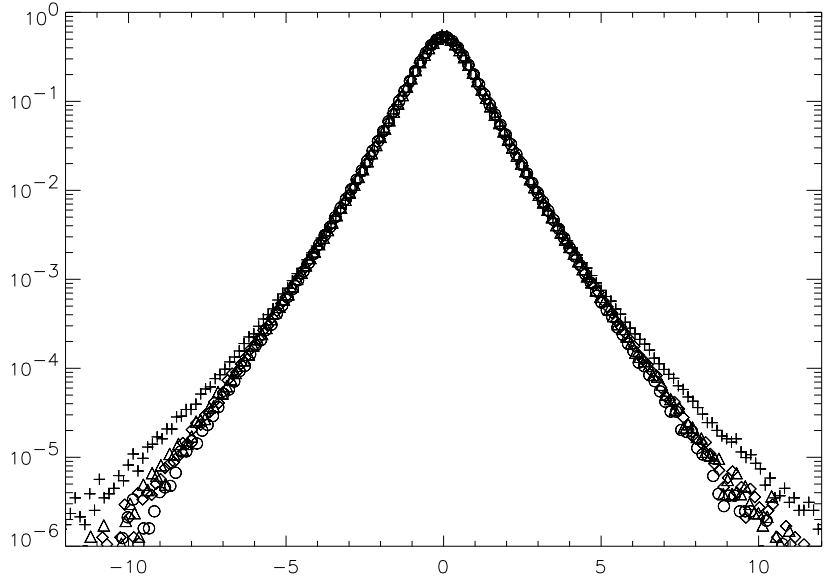,width=12cm}}
\caption[]{Comparison of the PDF of the transverse velocity 
increments defined by Eq. (\ref{increments}) for  $\ell=2\pi/256$ 
($10^8$ statistics  at $t=0.48$ for the velocity field from the four 
simulations: DNS (circles), RDT (crosses), RDT + constant viscosity (diamonds) and RDT + 
viscosity computed from RNG (triangles)}
\label{fig:pdft1}
\end{figure}

\begin{figure}[hhh]
\centerline{\psfig{file=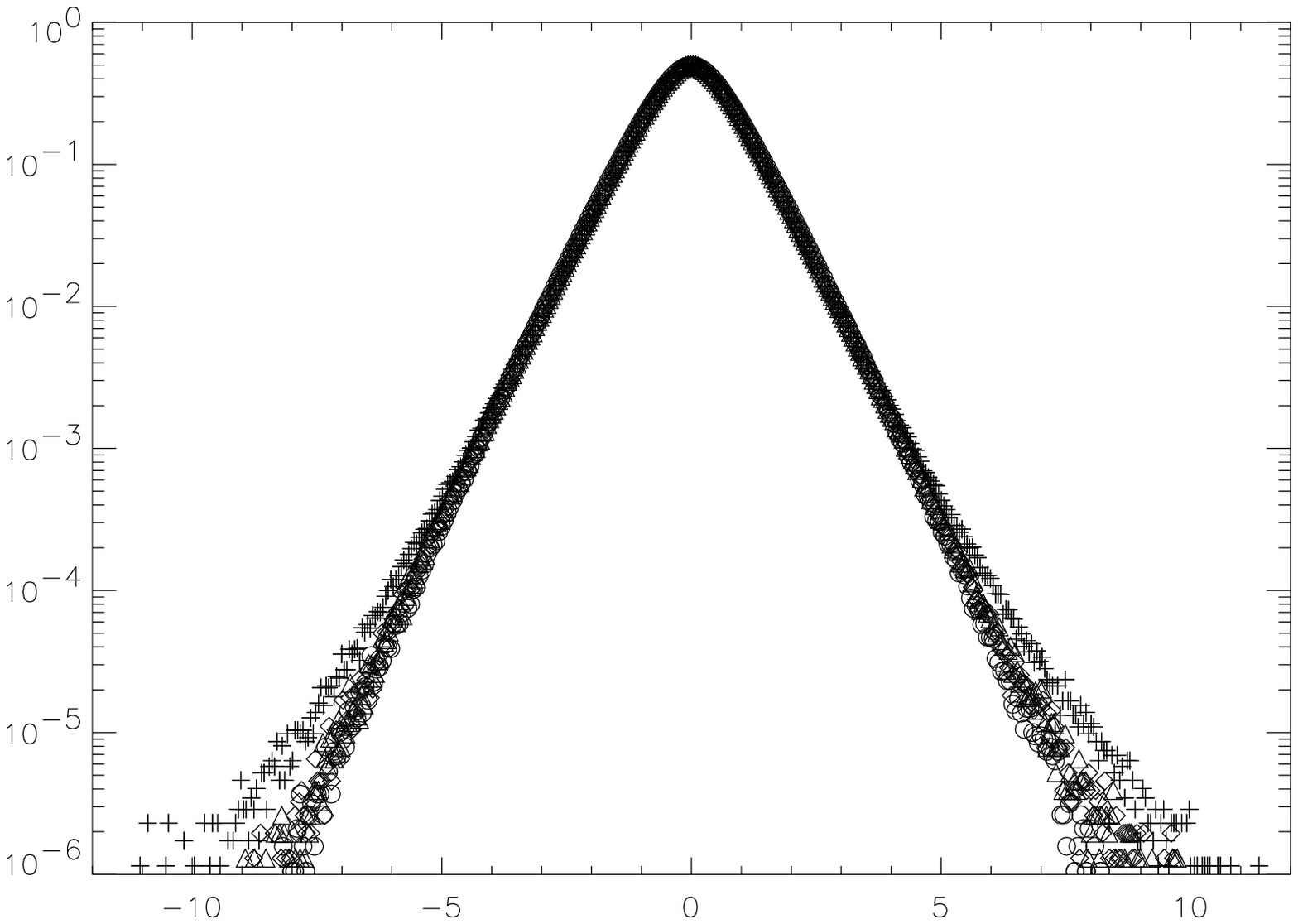,width=12cm}}
\caption[]{Same caption as in Fig. \ref{fig:pdft1} for $\ell=2\pi/64$}
\label{fig:pdft4}
\end{figure}

\begin{figure}[hhh]
\centerline{\psfig{file=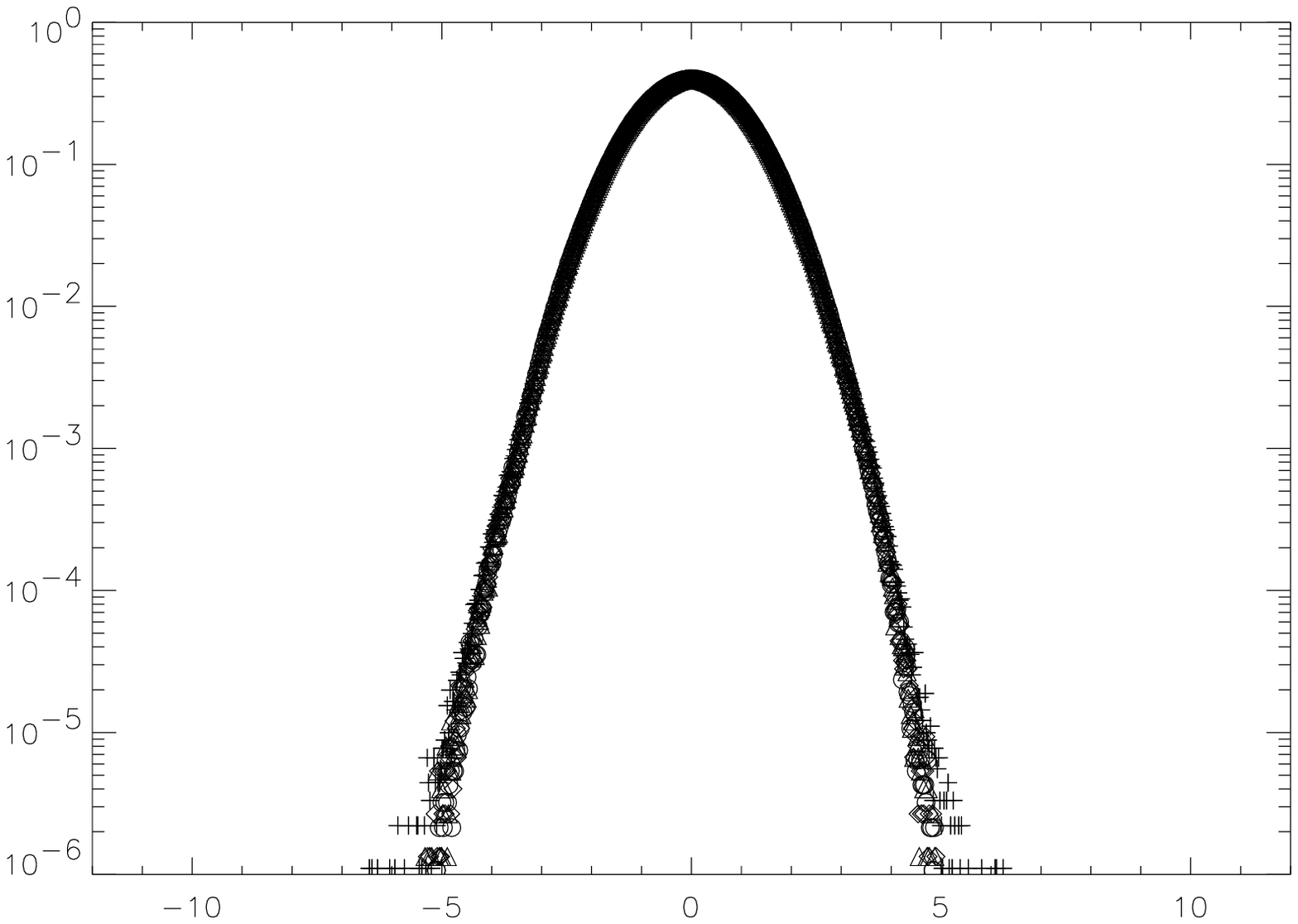,width=12cm}}
\caption[]{Same caption as in Fig. \ref{fig:pdft1} for $\ell=2\pi/4$}
\label{fig:pdft12}
\end{figure}

We conducted a statistical study of our velocity fields
corresponding to the end of the simulations ($t=0.48$). As usual
for study of the anomalous properties of turbulence, we consider
the velocity increments over a distance ${\bf l}$,
\EQ
\delta \bf u_{\bf l}={\bf u}({\bf x+{\bf l}})-{\bf u}({\bf x}).
\label{increments}
\EN
As usual, we will deal with the longitudinal and transverse to  ${\bf l}$
velocity increments, $\delta u_{l \parallel} = (\delta \bf u_{\bf l} \cdot {\bf l})/l$ and  $\delta u_{l \perp} =(\delta \bf u_{\bf l} \times {\bf l})/l$ respectively,
where  $l = |{\bf {\bf l}}|$. Figs. \ref{fig:pdfl1}, \ref{fig:pdfl4} and \ref{fig:pdfl12} show the probability distribution functions (PDF) of the longitudinal
 increments and  Figs. \ref{fig:pdft1}, \ref{fig:pdft4} and \ref{fig:pdft12} the PDF of transverse increments , for 3 values of $l$, 
obtained by the DNS, and our different RDT simulation (with and 
without turbulent viscosity).

 At large scale,
one observes a quasi-Gaussian behavior, with the development of
wider tails as one goes towards smaller, inertial scales.
This widening of the PDF's is a classical signature of the anomalous
scaling observed in turbulence. It can be measured by studying
the scaling properties of the velocity structure functions,
\EQ
S_p(\ell)=<\delta u_\ell^p>.
\label{structure}
\EN
In the inertial range, the structure function vary like $\ell^{\zeta_p}$.
For low Reynolds number turbulent flows, the scaling behavior in
the inertial range is very weak or undetectable because the inertial range
is very short.
\begin{figure}[hhh]
\centerline{\psfig{file=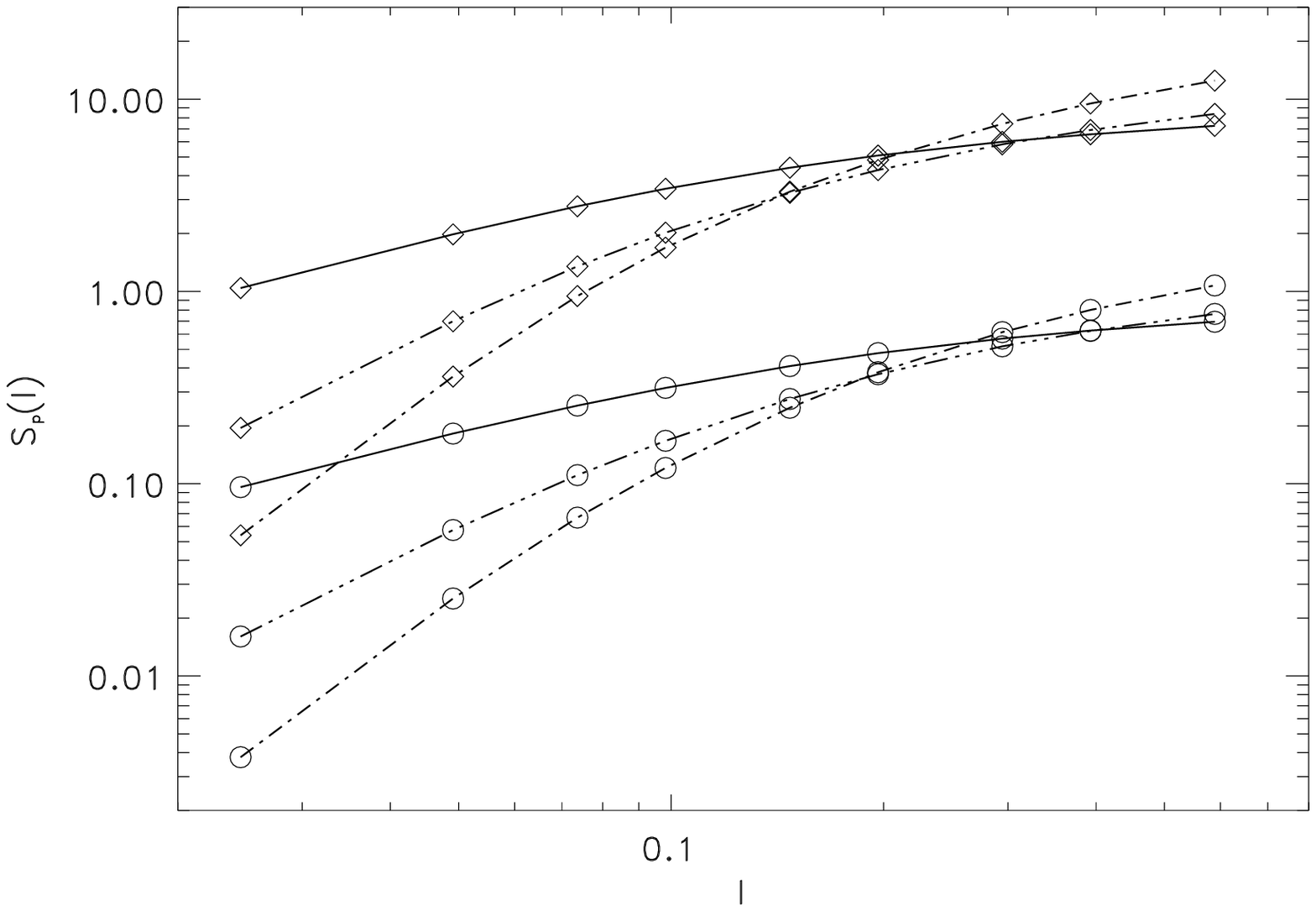,width=12cm}}
\caption[]{Structure functions computed with the PDF of the 
longitudinal velocity increments from DNS (circles) and the RDT 
(diamonds) for the first (solid line), second (dash-dotted line) and third 
(dash-dot-dot line) 
 moments (for convenience of presentation, the RDT structure 
functions have been multiplied by  10)}
\label{fig:structfctdnsrdtl}
\end{figure}
\begin{figure}[hhh]
\centerline{\psfig{file=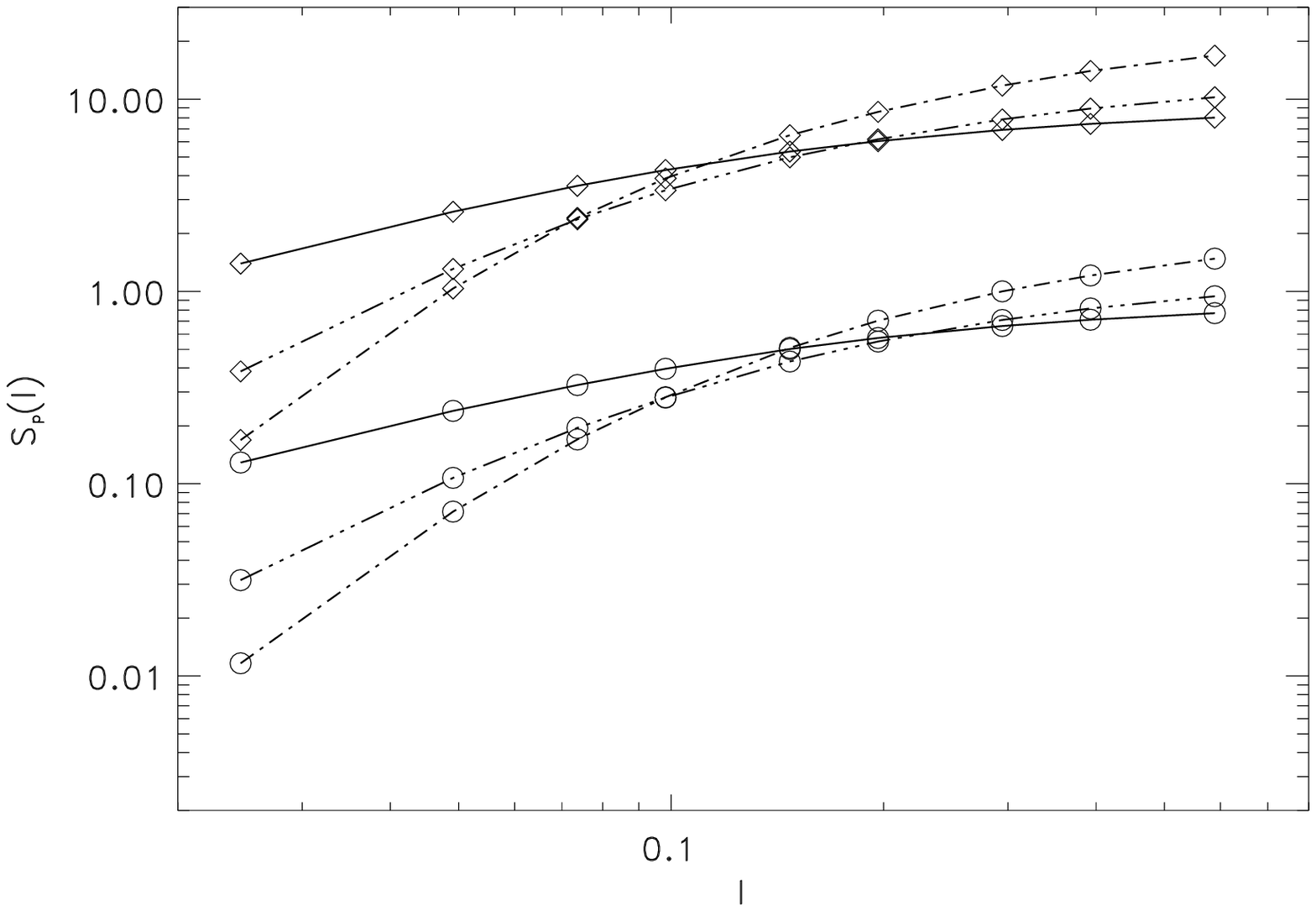,width=12cm}}
\caption[]{ Same plot as in 
the Fig. \ref{fig:structfctdnsrdtl} for the transverse velocity increments}
\label{fig:structfctdnsrdtt}
\end{figure}
 To exemplify this point, we show in Fig. \ref{fig:structfctdnsrdtl} 
 and \ref{fig:structfctdnsrdtt} the  structure functions  as a function of the scale separation for the longitudinal and the transverse velocity increments (the structure functions from the RDT simulation were shifted by a factor 10 for the clarity of the figures). Given the very weak scaling of our structure functions, we may use 
the extended self-similarity (ESS)
  property \cite{benzi93},
which states that $S_p(\ell)\sim S_3^{\zeta_p/\zeta_3}$ even outside
the inertial range of scales. We use this property because it allows to find
the scaling exponents in a more unambiguous way \cite{benzi93}.
\begin{figure}[hhh]
\psfig{file=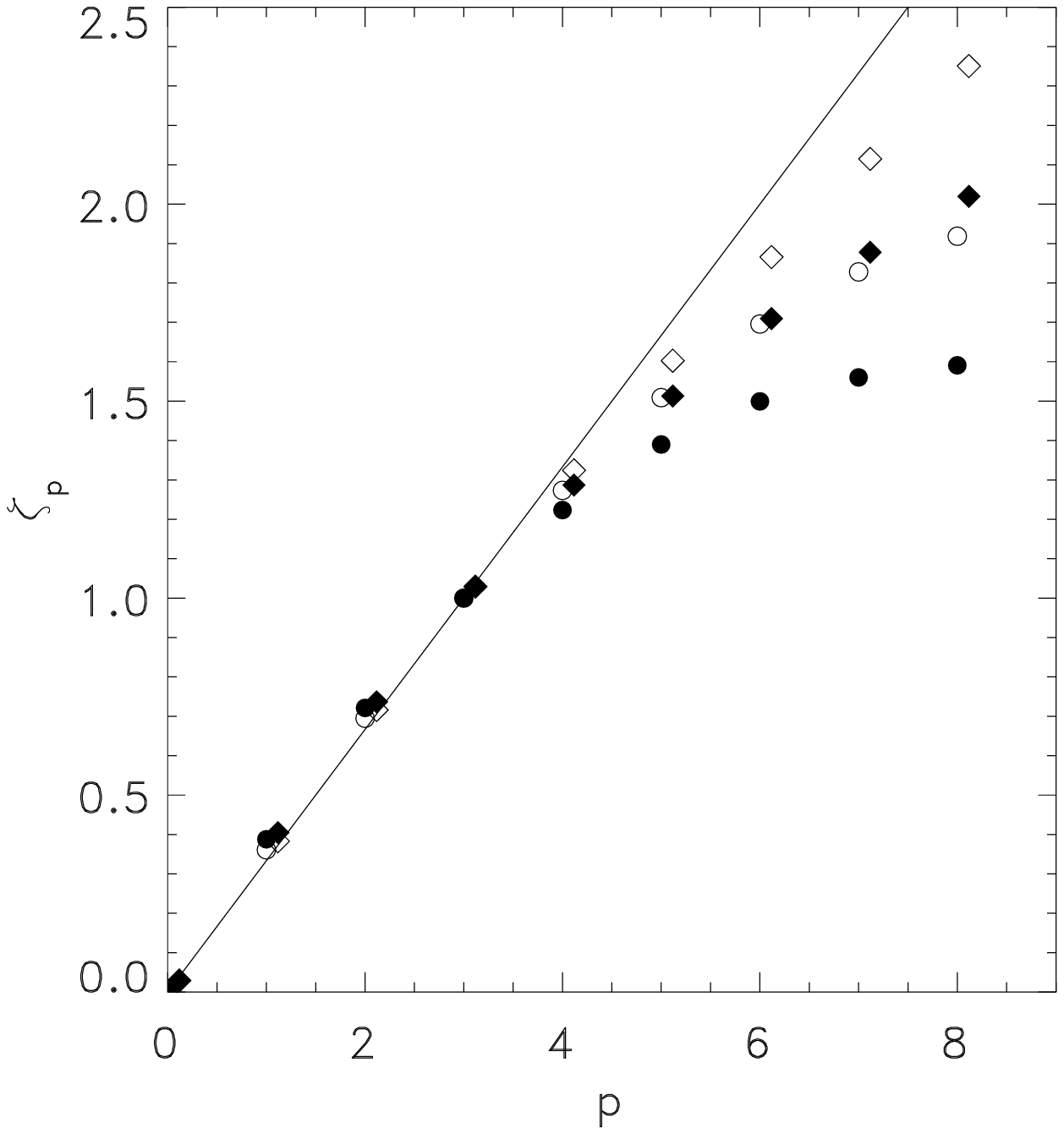,width=12cm}
\caption[]{Comparison of the scaling exponents computed from the DNS 
(diamonds) and the RDT (circles) statistics. The longitudinal exponents 
are plotted with empty symbols and the transverse exponents with 
filled symbols. }
\label{fig:scalingexpdnsrdt}
\end{figure}
\begin{table}
\begin{center}
\caption[]{scaling exponent of the velocity structure functions 
measured using the ESS property, in various simulations, at t=0.48.
\label{tab:scalingexp}}
\begin{tabular}{|c|c|c|c|c|c|c|}
\hline
\mbox{ }  &\multicolumn{6}{|c|}{longitudinal} \\
\hline
Order        &  K41  &  DNS   &  RDT  & RDT + Visc. 1 & RDT + Visc. 2 & local  \\
\hline
1            & 0.333 & 0.353  & 0.361 & 0.353         & 0.354 & 0.344 \\
2            & 0.667 & 0.687  & 0.695 & 0.687         & 0.687 & 0.678 \\
3            & 1.    & 1.     & 1.    & 1.            & 1.    & 1. \\
4            & 1.333 & 1.295  & 1.273 & 1.294         & 1.293 & 1.310 \\
5            & 1.667 & 1.573  & 1.509 & 1.570         & 1.568 & 1.607 \\
6            & 2.    & 1.836  & 1.696 & 1.830         & 1.824 & 1.892 \\
7            & 2.333 & 2.085  & 1.828 & 2.073         & 2.064 & 2.164 \\
8            & 2.667 & 2.321  & 1.919 & 2.301         & 2.287 & 2.423 \\
\hline
 \mbox{ } & \multicolumn{6}{|c|}{transverse} \\
\hline
Order        &  K41  &   DNS &  RDT  & RDT + Visc. 1 & RDT + Visc. 2 & local \\
\hline
1            & 0.333 & 0.375 & 0.388 & 0.376         & 0.376         & 0.352 \\
2            & 0.667 & 0.707 & 0.721 & 0.708         & 0.709         & 0.685 \\
3            & 1.    & 1.    & 1.    & 1.            & 1.            & 1.    \\
4            & 1.333 & 1.258 & 1.223 & 1.255         & 1.253         & 1.297 \\
5            & 1.667 & 1.484 & 1.390 & 1.477         & 1.468         & 1.578 \\
6            & 2.    & 1.680 & 1.499 & 1.664         & 1.645         & 1.843 \\
7            & 2.333 & 1.849 & 1.560 & 1.822         & 1.782         & 2.094 \\
8            & 2.667 & 1.991 & 1.591 & 1.952         & 1.882         & 2.333 \\
\hline
\end{tabular}
\end{center}
\end{table}
The measured exponents in our DNS are shown in
  Fig \ref{fig:scalingexpdnsrdt} and in Table \ref{tab:scalingexp}.
In both the longitudinal and transverse case, they are in agreement with the
previously reported exponents \cite{arneodo96}
  and they display a clear
deviation from the "non-intermittent" value $\zeta_p=p/3$.
However the difference between transverse and longitudinal exponents 
appear to be somewhat larger than the one observed  by 
Dhruva et al \cite{dhruva97} and Camussi et al \cite{camussi97}.
Corresponding quantities for the RDT simulation are shown
in  Fig \ref{fig:scalingexpdnsrdt} and Table \ref{tab:scalingexp}.
  One sees that the RDT statistics display larger and more
intermittent PDF tails for small scales, which makes the scaling
exponents to take smaller values corresponding to larger anomalous 
corrections. Again,
this situation is reminiscent of the case of the boundary layer.
In fact, the measured values in our simulation are remarkably similar
to those reported in the atmospheric boundary layer \cite{dhruva97}
or in a turbulent boundary layer \cite{stolovitsky93}. They are in between
the values two different values measured by Toschi (\cite{toschi99}) in
and above the logarithmic layer in numerical DNS of a channel flow.
A summary of these results is given in Table \ref{tab:scalingexpexp}
\begin{table}
\caption[]{scaling exponents of the velocity structure functions from : atmospheric turbulence at $10000<R_\lambda<15000$ (Dhruva) \cite{dhruva97}, channel flow (Toschi) \cite{toschi99} near the wall ($20<y^+<50$) and far from the wall ($y^+>100$) at $Re=3000$ and boundary layer at $R_\delta=32000$ (Zubair) \cite{stolovitsky93,zubair93}}
\begin{center}
\begin{tabular}{|c|c|c|c|c|c|}
\hline
  &\multicolumn{4}{|c|}{longitudinal} & transverse  \\
\hline
Order        & Dhruva &  Zubair &  Toschi $20<y^+<50$  & Toschi $y^+>100$ &  Dhruva \\
\hline
1            & 0.366 &  -     & 0.44                  & 0.37           & 0.359    \\
2            & 0.700 & 0.70   & 0.77                  & 0.70           & 0.680    \\
3            & 1.000 & 1.00   & 1.00                  & 1.00           & 0.960    \\
4            & 1.266 & 1.20   & 1.17                  & 1.28           & 1.200    \\
5            & 1.493 & 1.52   & 1.31                  & 1.54           & 1.402    \\
6            & 1.692 & 1.62   & 1.44                  & 1.78           & 1.567    \\
7            &  -    & 1.96   & 1.55                  & 2.00           & -        \\
\hline
\hline
\end{tabular}
\end{center}
\label{tab:scalingexpexp}
\end{table}
When a turbulent viscosity is added to the RDT simulation, the
intermittent wings are less pronounced in the PDF's
 and the anomalous correction decrease (Table \ref{tab:scalingexp}),
becoming similar to those observed in the DNS. This agrees with the
picture in which the anomalous corrections are determined by the non-local
interactions, while the local interactions act to restore the classical
Gaussian (Kolmogorov-like) behavior. Obviously, the shape of the 
turbulent viscosity also influences the intermittency correction: for 
the transverse case, where there is no asymmetry of the PDF's, both 
the constant  turbulent viscosity and the RNG turbulent viscosity 
provide intermittency corrections which are of the same level as the 
DNS. This is quite remarkable, since they include only one adjustable 
parameter, tuned as to conserve the total energy. For the 
longitudinal case, where an asymmetry is present, the two 
prescription give noticeably different result: as one goes towards 
lower scales, and as the asymmetry becomes larger between the 
positive and the negative increments, the PDF's computed  with RDT 
and constant turbulent viscosity display tails which are very close 
to that of DNS, while the PDF's of the RDT with turbulent RNG 
viscosity have a tendency towards a symmetrical shape, thereby 
failing to reproduce the DNS behavior. This difference of behavior 
between the RNG and constant turbulent viscosity will be further 
investigated in Section 2.\

\subsection{Comparison of the ``local'' experiment with DNS}

\begin{figure}[hhh]
\centerline{\psfig{file=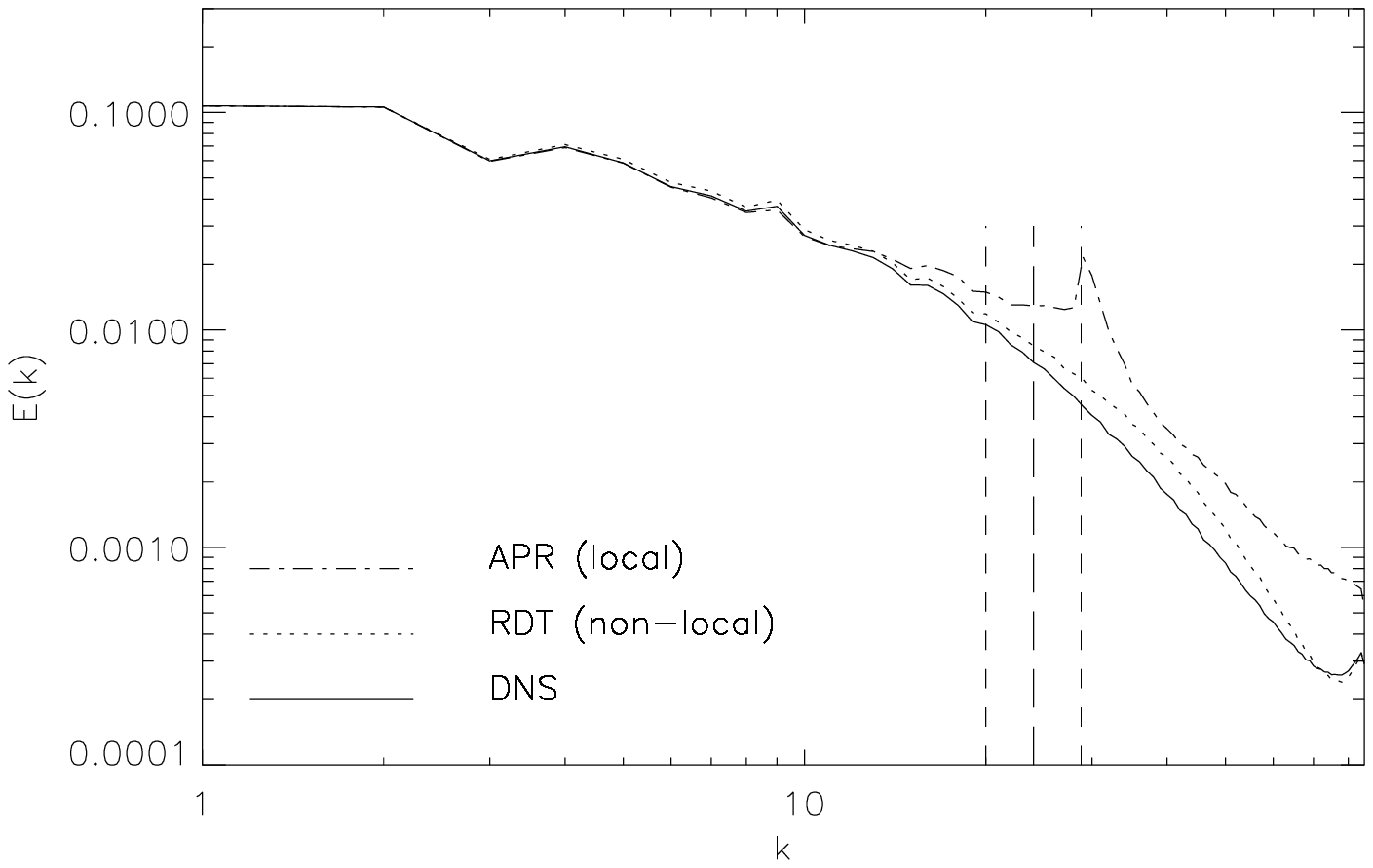,width=12cm}}
\caption[]{Comparison of energy spectra obtained via the DNS,  RDT and the 
``local'' simulation at t=0.48}
\label{fig:compengspectdnsrdtloc}
\end{figure}

Given the comparison of the DNS and the RDT (``non-local'') simulation,
one could  argue that the increase of the intermittency in  RDT 
is mostly due to the increased mean intensity of the small scales
(which is seen on the energy spectrum plot). A similar increase of
the small-scale intensity could be produced by other means which have
nothing to do with non-locality, e.g. by 
reducing viscosity in DNS. Will there be stronger intermittency in
all of such cases too? In order
to prove that it is not the case, we perform a simulation where, as the
  opposite of the RDT one, the non-local interactions at small scales were
  removed from the NS equation and only the local interactions were retained. In order to keep the local interactions which involve scales close to the cutoff
  scales, the velocity and the vorticity fields were split into three
  parts: the  large scales , the medium scales near the cutoff,
  and the small scales. This decomposition is defined in Fourier
space as follows,

\begin{eqnarray}
\uu(\kk)    &=& \uu_{ls}(\kk)    + \uu_{ms}(\kk)    + \uu_{ss}(\kk), \\
\omega(\kk) &=& \omega_{ls}(\kk) + \omega_{ms}(\kk) + \omega_{ss}(\kk)
\end{eqnarray}
where 
\begin{equation}
\begin{array}{lllll}
\uu_{ls}(\kk)  &=& \uu(\kk) &\mbox{for}&     k < k_c/C \\ 
\mbox{ }       &=& 0        &\mbox{for}&     k > k_c/C  \\
\uu_{ms}(\kk)  &=& \uu(\kk) &\mbox{for}& k_c/C < k < C \, k_c \\
\mbox{ }       &=& 0        &\mbox{for}& k_c/C < k \mbox{ and }   k > C \,k_c\\
\uu_{ss}(\kk)  &=& \uu(\kk) &\mbox{for}& C \, k_c < k  \\
\mbox{ }       &=& 0        &\mbox{for}&     k > C \,k_c \\
\end{array}
\end{equation}
Using these definition, the equation for the ``local'' simulation was 
the following:
\begin{eqnarray}
\partial_t \uu(\kk) &+& P(\uu \cdot \nabla \omega)(\kk) \nonumber \\
&-& \left[ P(\uu_{ls} \cdot \nabla \omega_{ss})(\kk) +
P(\uu_{ss} \cdot \nabla \omega_{ls})(\kk)\right]_{\{k>k_c\}}
= \nu \Delta \uu(\kk)
\end{eqnarray}
Where P is the projector operator. In our simulation we choose $C=1.2$
and the same cutoff scale $k_c=24$.  The result of this simulation are
  compared to the equivalent results from the DNS and the RDT simulation.
  The energy spectra are compared in Fig. \ref{fig:compengspectdnsrdtloc}.
\begin{figure}[hhh]
\centerline{\psfig{file=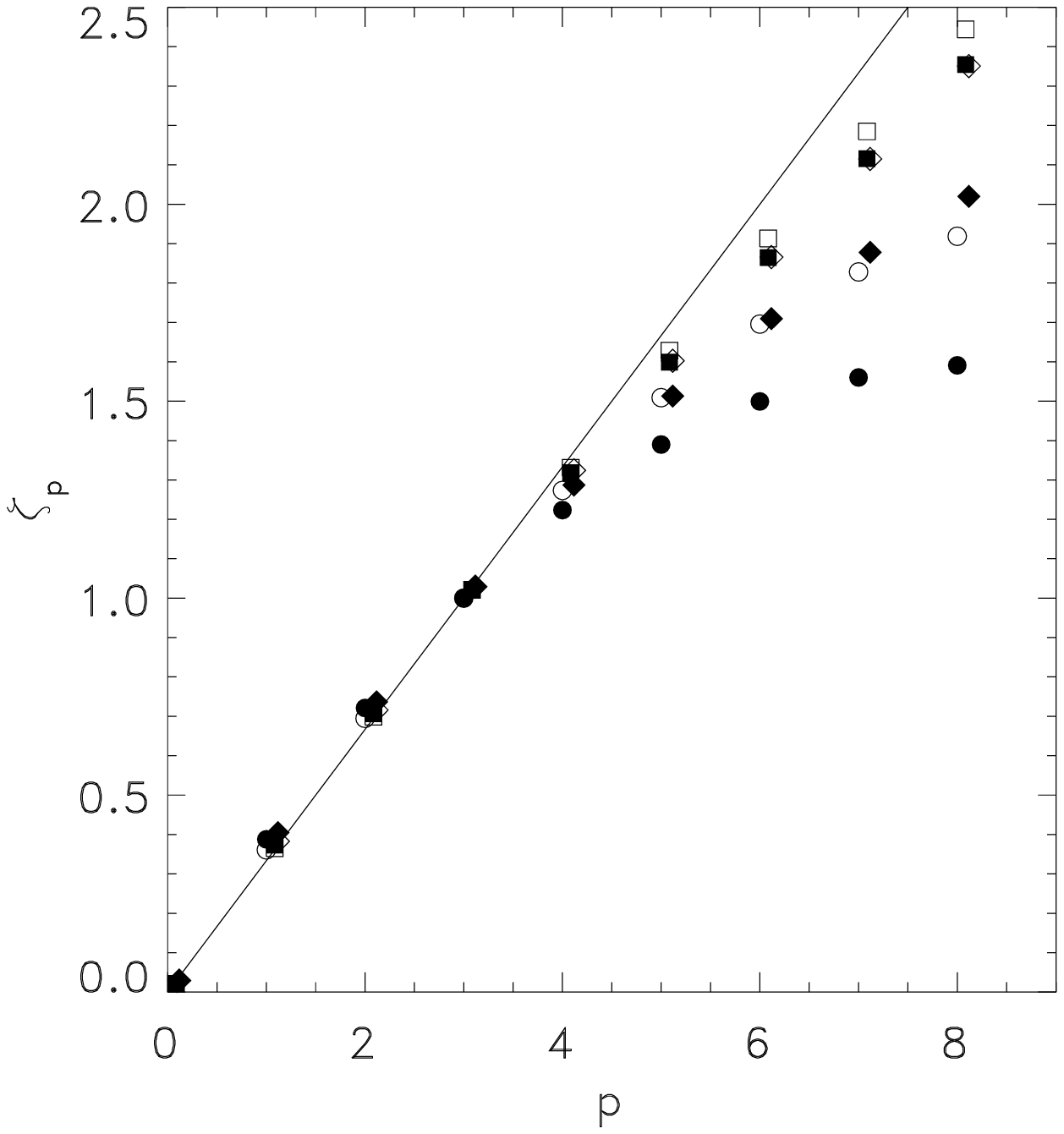,width=12cm}}
\caption[]{Comparison of the scaling exponents computed from the DNS 
(diamonds), RDT (circles) and the ``local'' simulation (squares) statistics. 
The longitudinal 
exponents are plotted with empty symbols and the transverse exponents with filled symbols.}
\label{fig:scalingexpdnsrdtloc}
\end{figure}
  This  ``local'' simulation contain more energy at small scales than
  the DNS even the RDT. The bump of energy near the cutoff scale $k_c$ is
  due to the fact that the ``local'' approximation is introduced only for scales
  smaller than $k_c$. Despite the high level of energy at small scales,
  the solution of this ``local'' simulation is much less intermittent
  than the equivalent field from DNS and RDT. A comparison of the scaling exponents
is shown Fig. \ref{fig:scalingexpdnsrdtloc} and table \ref{tab:scalingexp}
  for both the longitudinal and transverse velocity increments. These results confirm the
  idea that intermittency is caused by the non-local interactions and not just
by mere presence of the small scales.

\clearpage

\section{Qualitative explanation of the intermittency}

Our results can be used to get a qualitative understanding of the 
intermittency via  the scale behavior of both the PDF of the velocity 
increments and its moments. For this, we are going to build a toy 
model of turbulence, mimicking the small scale non-local dynamics. In 
spirit, this amounts to an "anti-shell" model of turbulence, because, 
here, we retain only interactions between distant wave-numbers, by 
contrast with the ordinary shell model 
\cite{gledzer73,yamada88,yamada89} which theoretically only retains 
local interactions. In this context, it is interesting to note that 
the ordinary shell model requires a certain degree of non-locality 
between modes so as to generate intermittency: it can indeed be 
proven that the intermittency correction disappear when the 
separation between two consecutive shell tends to zero 
\cite{dombre00}. Another known pitfall of the shell model is its 
incapacity to describe the observed skewness (asymmetry) generation 
along the scale of the PDF of the longitudinal increments. This is annoying, since this 
skewness  is directly related to the non-zero value of the third 
order moment, and, hence, to the essence of the Kolmogorov cascade 
picture via the 4/5 law.
Finally, the original shell
model is very crude, since there is no spatial structure (everything 
is described by
Fourier modes). We now show how elaborate a cleaner toy model
of turbulence using localized wave-packets, leading to a description
of the small-scale statistics in term of Langevin processes subject 
to coupled multiplicative and additive noise.

\subsection{The Langevin model of turbulence}

Our numerical simulations showed that both the energy spectrum (and 
decay) and the
intermittency quantities are well reproduces by a model in which only 
the non-local
interactions are left in the small-scale equations whereas the local 
interactions
of small scales among themselves are replaced by a turbulent viscosity term.
Such a model is described by equations  (\ref{eq:seteqs}) with $\nu$ replaced
by the turbulent viscosity coefficient $\nu_t$ in the small-scale equations,
\EQA
\partial_t u_i +\partial_j (U_i u_j)+\partial_j (u_i U_j)
&=& -\partial_i p +\nu_t \Delta u_i+\sigma_i, \label{eq:seteqs2} \\
\partial_j u_j &=&  0 \nonumber,
\ENA
and $\sigma_i$ is given by (\ref{pseudoforce}).
We are interested
in the contribution of non-local interactions to the statistics of the
non-Gaussian small scales. For this, we assume that the large scale $L$
quantities ( ${\bf U}$  and its derivative, and $\sigma_i$) are fixed 
external processes, with prescribed
statistics (to be defined later), and derive an equation for the 
small scale ${\bf l}$ velocity field
${\bf u'}$,
by taking into account the scale separation $\ell/L=\epsilon\ll 1$.
For this, we decompose the velocity field into localized wave-packets
via a Gabor transform (GT) (see \cite{nazarenko99d})
\EQ
{\hat u}({\bf x},{\bf k},t)=\int g({\epsilon_\ast \vert
\bf x-x'}\vert)e^{i{\bf k\cdot (x-x')}}
{\bf u}({\bf x'},t)d{\bf x'},
\label{gabordef}
\EN
where $g$ is a function which decreases rapidly at infinity and
$1\ll\epsilon_\ast\ll\epsilon$. Note that the GT of ${\bf u}$
is a natural quantity for the description of the velocity
increments because of the following relation
\EQ
{\bf u}\left({\bf x}+{\bf {\bf l}}\right)-
{\bf u}\left({\bf x}-{\bf {\bf l}}\right) = {1 \over 2i}
\int e^{-{\bf {\bf l}}\cdot{\bf k}} Im \, {\bf {\hat u}}
({\bf x},{\bf k}) \, dk.
\EN
Thus,  velocity increments are related to GT via the Fourier transform,
and all information about the $l$-dependence is contained in the GT
dependence on $k$ (the main contribution to the above integral
comes from $k \sim 2\pi/l$.). On purely dimensional ground, 
we see that $\hat{u} \sim k^{-d}\delta u$, where $d$ is the dimension. 
Therefore, in the sequel, we shall identify $k^d Im \, \hat{u}$ with the 
velocity increment.

Applying GT to (\ref{eq:seteqs}) we have (see \cite{nazarenko99} for details):
\begin{equation}
D_t \hat{{\bf u}}  =
\hat{{\bf u}} \cdot \hat {\bf \xi}+\hat
\sigma_\perp - \nu_t k^2 \hat{{\bf u}}, \label{smallscales}
\end{equation}
where $\xi$ and $\sigma_\perp$ are random processes, given by
\EQA
\hat{\bf \xi}&=&\nabla
\left(2\frac{\bf k}{k^2}{\bf U}\cdot {\bf k}-{\bf U}\right),\nonumber\\
\hat{\sigma}_\perp &=& \hat{\sigma} - {{\bf k}\over k^2}
({\bf k}\cdot\hat{{\bf \sigma}}),
\label{noiseses}
\ENA
and
\begin{eqnarray}
D_t & = & \partial_t + \dot{{\bf x}}\cdot\nabla + \dot{{\bf 
k}}\cdot\nabla_{{\bf k}},
\nonumber \\
\dot{{\bf x}}  & = &  {\bf U} =  \nabla_{k} H, \label{ray11}\\
\dot{{\bf k}}  & = & - \nabla ({\bf k}\cdot{\bf U}) = - \nabla H,
\label{ray22} \\
H & = & {\bf U}\cdot{\bf k}.
\label{frequency}
\end{eqnarray}
Because the large-scale dynamics  is local in $k$-space, it is only
weakly affected by the small scales and the quantities ${\bf \xi}$ and
${\bf \sigma}$ in equation (\ref{smallscales})
  can be considered as a given noise. Also, because the equation is 
linear in $\hat{u}$, we immediately see that $k^d \hat{u}$ will also 
satisfy an equation similar to (\ref{smallscales}) subject to a 
straightforward modification of the force definitions. Before 
elaborating more on (\ref{smallscales}), it is convenient to study in 
closer details the physical parameters of this equation.

\subsection{The noises}

\begin{figure}[hhh]
\centerline{\psfig{file=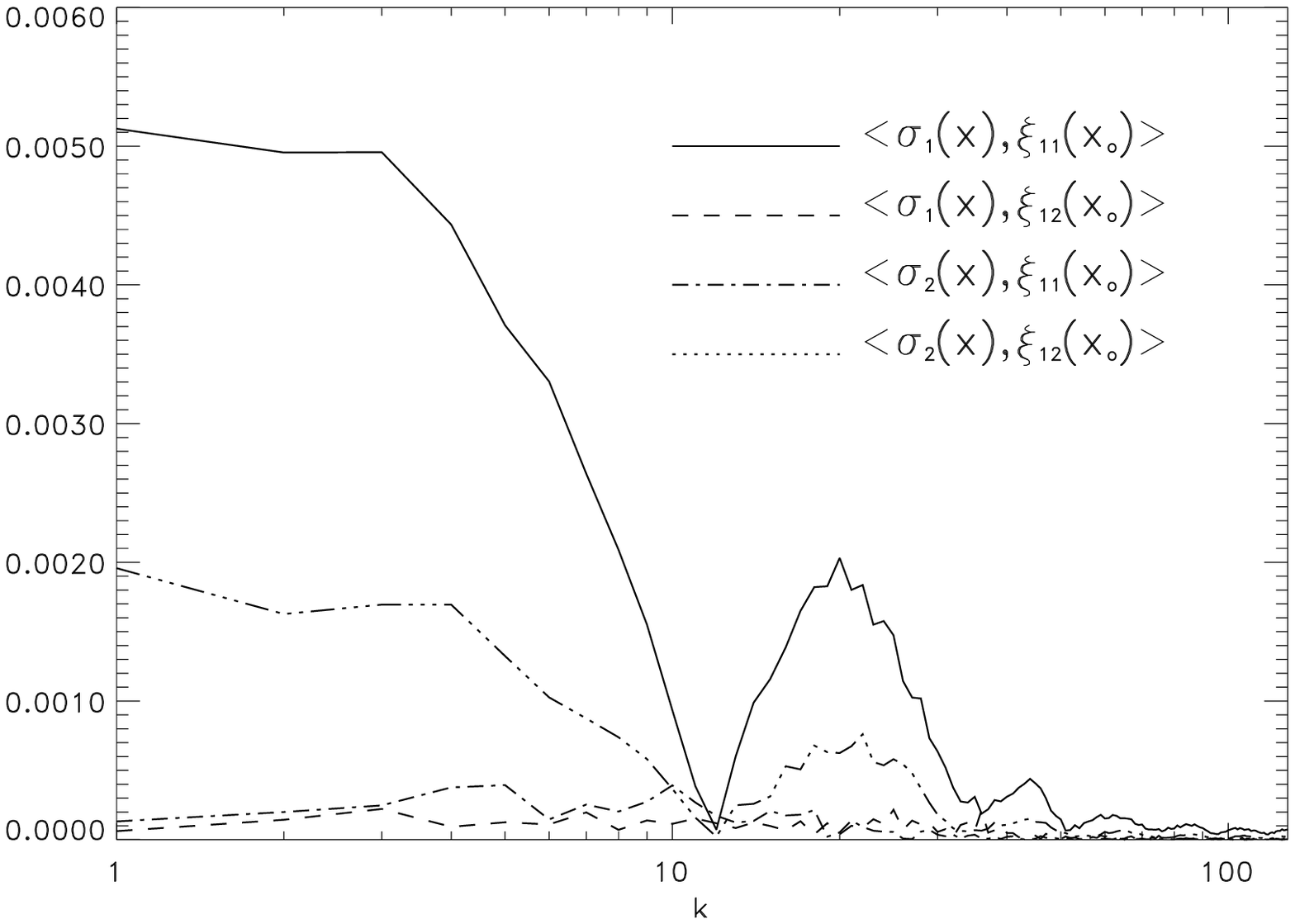,width=12cm}}
\caption[]{Fourier transform of the space cross correlation between 
the force $\sigma$ and the force $\xi$. Coordinates $y$ and $z$
are fixed.}
\label{fig:corel-ij-x}
\end{figure}

In equation (\ref{smallscales}), the noise $\xi$ and $\sigma_\perp$ 
appearing as the projector of two 
quantities, one related to the velocity derivative
 tensor, and another related to the 
Gabor transform of the energy transfer from large to small scales. 
In the sequel, we present 
a statistical study of these two noises in the physical space
(i.e. after inverse Gabor transforming $\hat \sigma_\perp$ ).
We will also consider the Fourier spectra of the corresponding
two-point correlations. Let us choose ${\bf k}$ 
to be along one of the coordinate axis (without loss of generality
because of the isotropy). Then the components of $\xi$
coincide (up to the sign) with the corresponding components of
the  velocity derivative tensor, and we will use this fact in the
rest of this section. The velocity 
derivative tensor  has been studied in the literature
e.g. in \cite{tsinober92,andreotti97} in terms of correlations 
between the directions of the anti-symmetrical part of the tensor 
(the vorticity) and the symmetrical part (the strain). Over long 
time, the vorticity appears to be aligned with the direction of 
largest stretching. Other studies focused on the PDF of the modulus 
of one component. For example, Marcq and Naert \cite{marcq00} observe 
that the derivative has a highly non-Gaussian distribution, but with 
a correlation function which decays rapidly, and can be approximated 
by a delta function at scales large compared to the dissipative 
scale. In the present case, we observe different features.   Because 
of the isotropy, we can concentrate only on two quantities, say 
$\xi_{11}$ and $\xi_{12}$.
\begin{figure}[hhh]
\centerline{\psfig{file=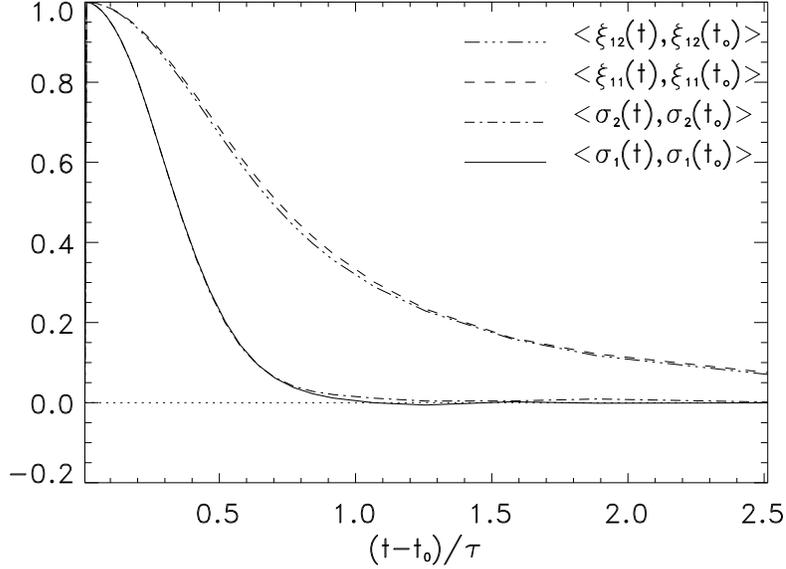,width=12cm}}
\caption[]{The normalized 
auto-correlation in time for the two force 
components  $\sigma$ and $\xi$ (the turnover time $\tau$ is equal to 0.19)}
\label{fig:corel-ii-t}
\end{figure}
 Fig. \ref{fig:corel-ii-t} shows the 
equal-position, time correlation 
$C_{1i}(t-t_0)=<\xi_{1i}(t)\xi_{1i}(t_0)>$ and $\alpha_{ii}(t-t_0) 
=<\sigma_{i}(t)\sigma_{i}(t_0)>$ as a function of 
$t-t_0$. Note that these quantities are normalized to 1
at $t=t_0$ in this figure. Firstly, we see that $C_{11}$ approximately
coincide with 
$C_{12}$ and $\alpha_{11}$ coincides with $\alpha_{22}$ which
is a good indicator of isotropy (without normalization 
there would be $C_{12} = -3 \, C_{11}$).
Secondly, we see that the correlations  $C_{11}$ and
$C_{12}$ decay to zero over a 
time scale which is of the order of few turnover times $\tau$ ($\tau=0.19$). 
On the other hand,  the correlations of $\sigma$ decay much faster, over a 
time of the order of $\tau/2$.
\begin{figure}[hhh]
\centerline{\psfig{file=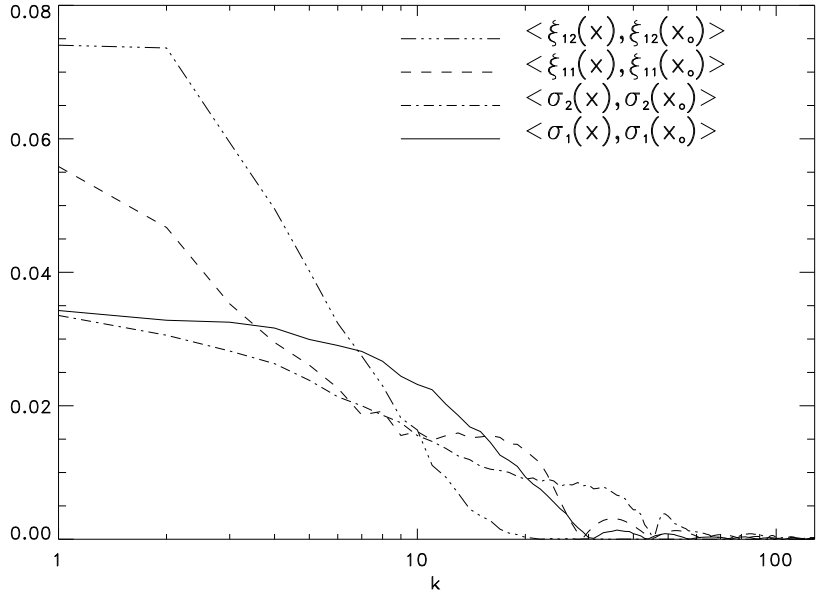,width=12cm}}
\caption[]{Fourier transform of the space auto-correlation for the two 
components of the force $\sigma$ and $\xi$. Coordinates $y$ and $z$
are fixed.}
\label{fig:corel-ii-x}
\end{figure}
 Fig. \ref{fig:corel-ii-x} displays the 
Fourier transforms of the equal-time two-point correlations
$D_{1i}(x-x_0)=<\xi_{1i}(x,y,z, t)\xi_{1i}(x_0,y,z,t)>$ 
and $\alpha_{ii}(x-x_0) 
=<\sigma_{i}(x,t)\sigma_{i}(x_0,t)>$. One sees that all correlations 
are very weak beyond $2k_c$. The correlation $D_{12}$ appears to be 
the largest at large scales, but it decays more rapidly than the other 
correlations.\
\begin{figure}[hhh]
\centerline{\psfig{file=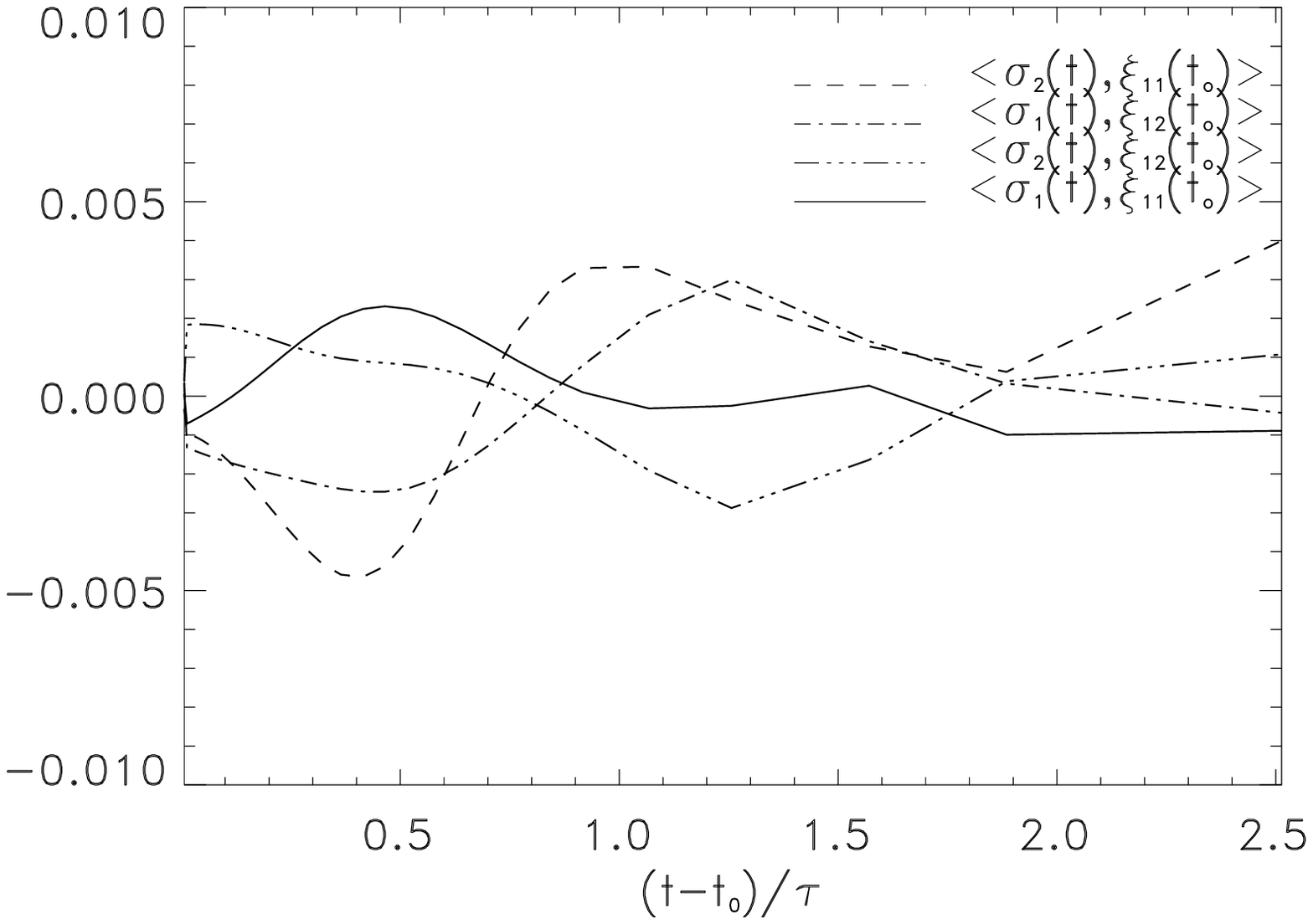,width=12cm}}
\caption[]{The cross-correlation in time between the forces $\sigma$ and 
$\xi$ ($\tau = 0.19$)}
\label{fig:corel-ij-t}
\end{figure}
We have also investigated the cross-correlations between the noises. 
The equal position cross-correlation are displayed in Fig. 
\ref{fig:corel-ij-t}. The correlation is rather weak, but there is a 
tendency for $\sigma_i$ to be correlated with $\xi_{1i}$ over a 
time scale of the order of $\tau$, while it is anti-correlated with 
the other component of the tensor, over such a time scale. The equal 
time correlations are shown in Fig. \ref{fig:corel-ij-x}.
 Notice that the cross correlation are one order of magnitude weaker than the 
direct correlations. The cross-correlation involving $\xi_{12}$ are 
essentially zero, while the correlations involving $\xi_{11}$ display 
a first overall decay up to $k=k_c$, followed by an extra bump up to 
the end of the inertial range ($k=40$). In the next Section, we will 
show that this feature is actually related to the energy cascade.\

One may also note that the two noises are spatially very intermittent.
\begin{figure}[hhh]
\centerline{\psfig{file=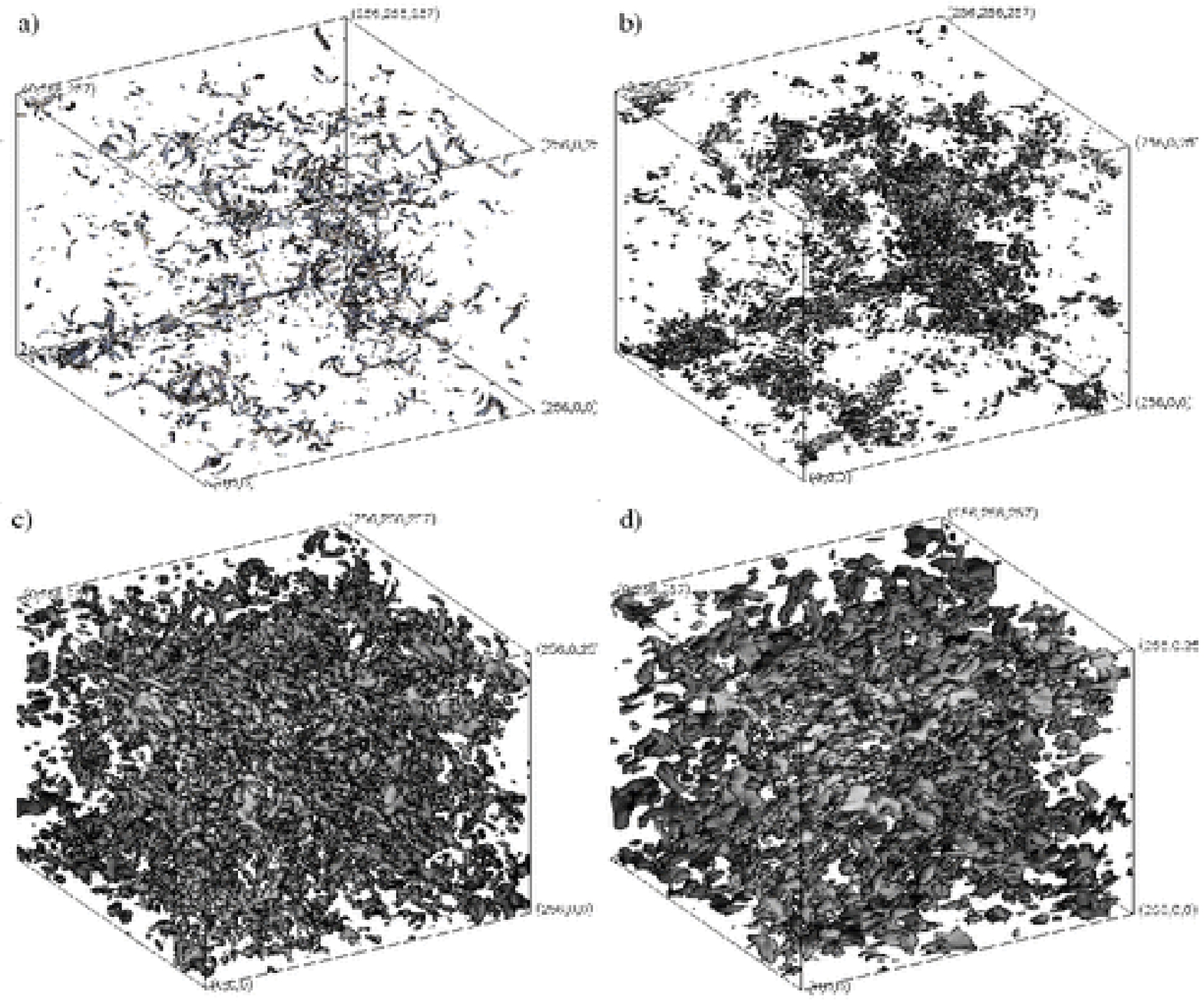,angle=-90,width=12cm}}
\caption[]{Isovalue (3.5 times the RMS) at $t_0 = 0$ of {\bf 
a)} the absolute value of vorticity, {\bf b)} the 
corresponding additive force ($\vert \sigma \vert$) {\bf c)} 
$\xi_{11} = \partial U_x / \partial x$ and {\bf d)} 
$\xi_{12} = \partial U_x / \partial y$ }
\label{fig:isoforc}
\end{figure}
In Fig. \ref{fig:isoforc}, we show iso-surfaces  of 
the modulus of the noises, corresponding to 3.5 times the RMS value. For 
comparison, the same plot is made for the vorticity. One observes 
well defined patches of $\sigma$ which are strongly correlated 
with areas of strong vorticity. In the case of $\xi$, the patches are 
much more space filling. The longitudinal component $\xi_{11}$ is 
characterized by smaller-scale structures than the transverse 
component $\xi_{12}$.\

\begin{figure}[hhh]
\centerline{\psfig{file=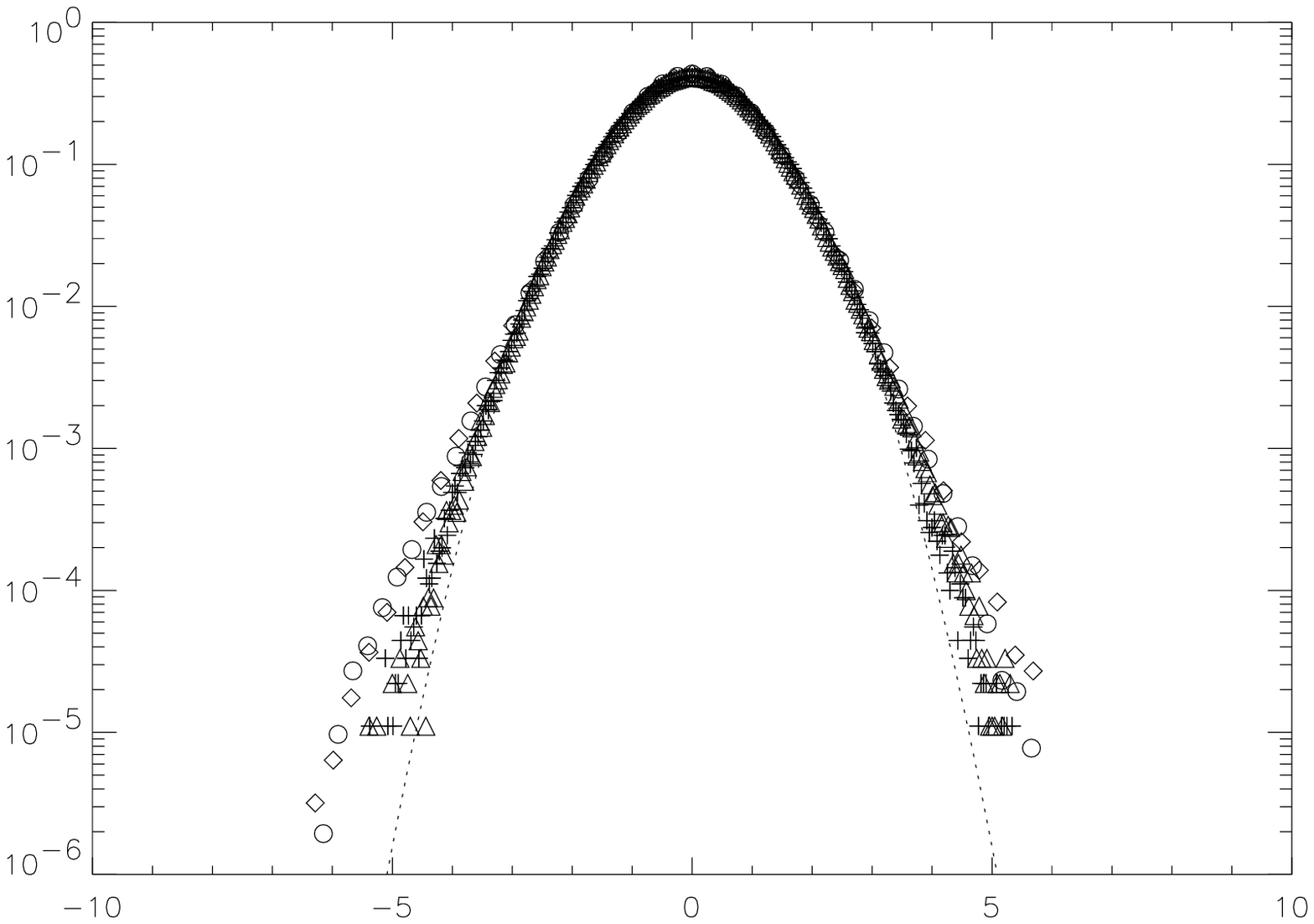,width=12cm}}
\caption[]{PDF of the  increments  $\xi_{11}(x+\ell_1,y,z) -
 \xi_{11}(x,y,z)$ for  $\ell_1= 2 \pi/256$ (circles) and $\ell_1= 2 \pi/4$ (crosses) and 
 $\xi_{11}(x,y+\ell_2,z) - \xi_{11}(x,y,z)$ for  $\ell_2= 2 \pi/256$ 
(triangles) and $\ell_2= 2 \pi/4$ (diamonds) (the dotted line corresponds to Gaussian statistics). } 
\label{fig:pdfi22}
\end{figure}

To obtain an indication about the scale variation 
of the statistical properties of the noises, we also computed the 
PDF's of the noise spatial increments
$\delta{\sigma_i}_{\ell}=\sigma(x+\ell)-\sigma(x)$ and 
$\delta {\xi_{ij}}_{\ell}=\xi_{ij}(x+\ell)-\xi_{ij}(x)$. 
Note that the first of these quantities is directly related with
the Gabor transform of the additive noise, whereas the second one
contains some useful information about the time correlations
via the Taylor hypothesis (which is valid locally because the
large-scale velocity is typically greater than the small-scale one). 
\begin{figure}
\centerline{\psfig{file=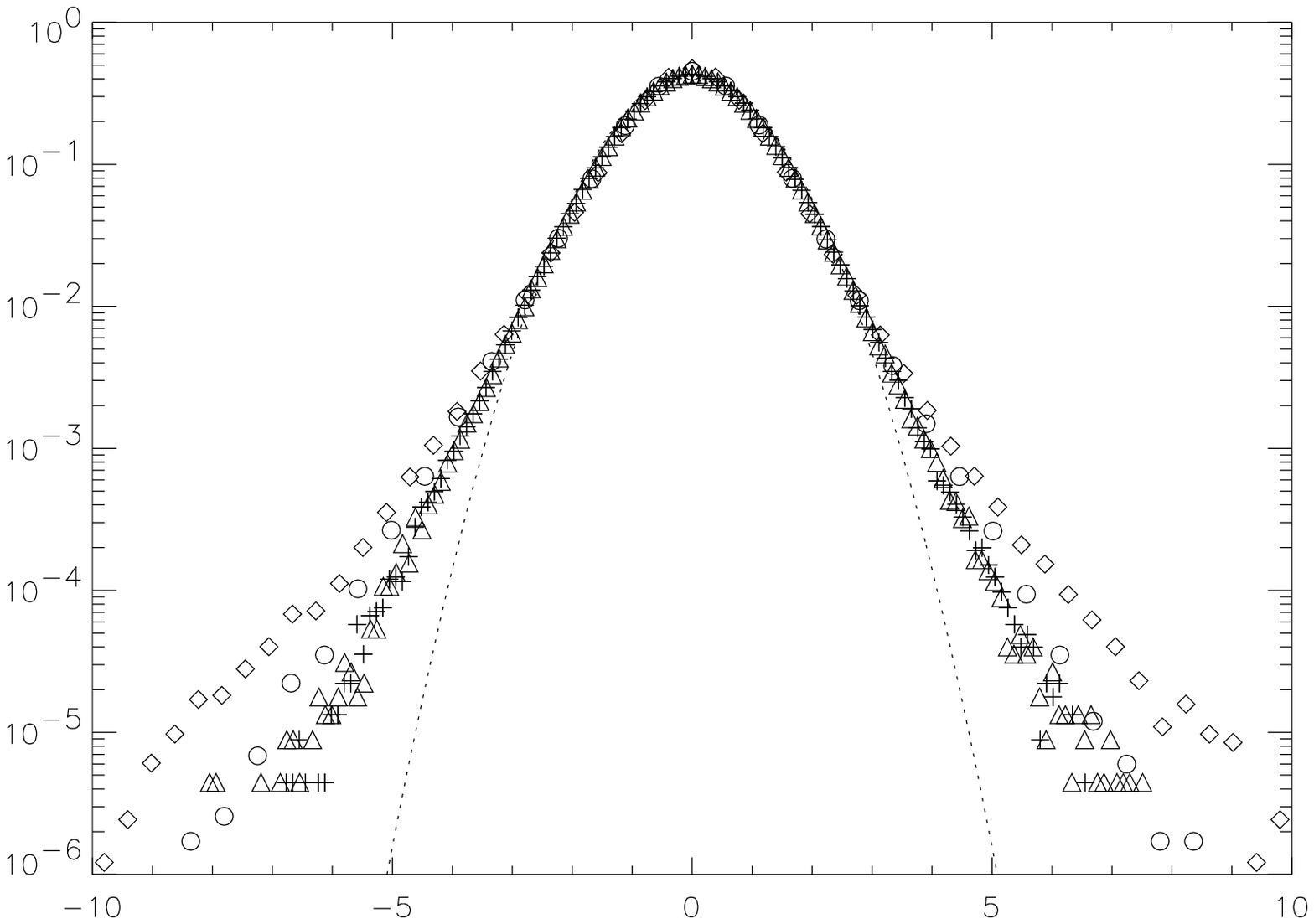,width=12cm}}
\caption[]{PDF of the increments  $\sigma_1(x+\ell_1,y,z)
 - \sigma_1(x,y,z)$ for  $\ell_1= 2 \pi/256$ (circles)
 and $\ell_1= 2 \pi/4$ (crosses) and  $\sigma_1(x,y+\ell_2,z) -
 \sigma_1(x,y,z)$ for  $\ell_2= 2 \pi/256$ (triangles)
 and $\ell_2= 2 \pi/4$ (diamonds) (the dotted
 line corresponds to Gaussian statistics).  }
\label{fig:pdfi11}
\end{figure}
Fig. \ref{fig:pdfi11} shows the results of the longitudinal and 
transverse increments for the first component
 of the additive noise $\sigma_1$ and the equivalent results for the 
component $\xi_{11}$ of the multiplicative multiplicative noise are shown 
on Fig. \ref{fig:pdfi22}.
 The PDF are displayed for $\ell= 2 \pi/256$ 
and $2 \pi/4$. One observes wide, 
quasi-algebraic tails for the additive noise, similar to those observed 
for the PDF's of the velocity derivatives. The PDF of $\xi_{11}$ are much 
closer to Gaussian statistic.

\subsection{The turbulent viscosity}
\label{sec:turbvisc}
In the previous Section, we have discussed the influence of two 
prescription for the turbulent viscosity, one based on the RNG, one 
taken simply as constant. In the sequel, we shall use the simple 
formula:\
\EQ
\nu_t=\left(\nu_0^2+B^2\left(\frac{u}{k}\right)^2\right)^{1/2} ,
\label{viscodimen}
\EN
where $\nu_0$ and $B$ are a constant and $u=k^d \, Im \, {\hat u}$ 
is the velocity 
increment over a distance $1/k$ (hereafter we drop $\delta$
in $\delta u$ ). When $B=0$, this formula provides 
the constant turbulent viscosity. When $\nu_0=0$, it provides a 
dimensional analog of the  RNG viscosity, and tends to zero as $k$ tends 
to infinity.

\subsection{Statistical properties of the velocity increments}

We are now going to derive qualitative results by adopting two 
complementary points of view: in a first one, we will study 
the statistical properties of the velocity increments in the
frame of reference moving together with the wave packets 
in $(k, x)$ space. This corresponds to
a  Lagrangian description in the scale space. In the second 
approach, we replace time with its expression in terms of $k$,
as it would follow from the ray equation
(\ref{ray22}). This will give as an
equation at a fixed $k$ which corresponds to an Eulerian description.
As a further 
simplification, we shall leave for further study the possible 
correlation between longitudinal and transverse velocity increments 
described for example in \cite{andreotti97t} and consider a
one-dimensional version of (\ref{smallscales}), treating
  the quantity $u=k \, Im \, \hat{u}$ as a ``velocity increment'' over
the distance $l= 2 \pi/k$,
\begin{eqnarray}
D_t u &=& u \xi + \sigma_\perp - \nu_t k^2 u, \label{ray221D} \\
\dot{{k}}  & = & - {k} \xi
\label{ray221Da} 
\end{eqnarray}
Here, we assumed that the forcing to be symmetric such that
it does not produce any $Re \, \hat{u}$.
 This  toy model can also be viewed as
a passive scalar in a compressible 1D flow. Artificial introduction
of compressibility is aimed at modeling the RDT stretching effect
which appears only in the higher number of dimensions for incompressible
fluids. 

Study of the noises  in Section 2.2 
revealed their rich and complex behavior.
As a first simplifying step, we disregard these complexities and
  use the Gaussian, delta correlated approximation, as will be
done in the next two sub-sections.
Given a rather short time correlation of $\sigma$, our delta 
approximation is rather safe. The delta approximation for $\xi$ is 
more debatable, and the performance of such  a model should be 
further examined in future. 
Also, the Gaussian hypothesis is obviously only valid at large scale, 
and for $\xi_{11}$. Therefore, the generalization of ours results for 
non-Gaussian noises would be very interesting, and is the subject of 
an ongoing research. In the sequel, we consider the function 
$\alpha$, $D$ and $\lambda$ as free parameters.

\subsubsection{The Lagrangian description}

In  the frame of reference moving  with the wave packets 
in $(k, x)$ space, the l.h.s. of (\ref{ray221D})  becomes
simply the time derivative. On the other hand, $k$ has to be
replaced in terms of its initial value $k_0$ and time everywhere
including the noises $\sigma$ and $\xi$. Such a transformation
from the laboratory to the moving frame can obviously change the
statistics of $\sigma$ and $\xi$. In the Lagrangian description
we will assume that we deal with noises which are Gaussian 
in the moving frame  with correlations functions 
\EQA
<\sigma_{\perp}(k(t),t)\sigma_{\perp 
}(k'(t),t')>&=&2\alpha \, \delta(t-t'),\nonumber\\
<\xi(k(t),t)\xi(k'(t),t')>&=&2D \,  \delta(t-t'),\nonumber\\
<\xi(k(t),t)\sigma_{\perp }(k'(t),t')>&=&2\lambda \, \delta(t-t'),
\label{correforces}
\ENA
where coefficients $\alpha, D$ and $\lambda$ depend on the scale
via $k_0$. With these noises, (\ref{ray221D})  becomes
a Langevin  equation for the
  velocity increments, where $\xi$ is a multiplicative noise, 
$\sigma_\perp$ is an additive noise, and $\nu_t k^2 u$ the ( 
non-linear) friction.
The multiplicative noise is produced by interaction of
two small scales with one large scale whereas the additive noise
is due to a merger of
two large scales with into one  small scale (therefore, the later
acts at the largest among the small scales). For Gaussian, delta 
correlated noises, this Langevin equation leads to a Fokker-Planck 
equation for the probability distribution $P_k(u,t)$ of the velocity 
increment $ u$, 
\EQA
\partial_t P_k&=&\partial_u \left(\nu_t k^2 u P_k\right)\nonumber\\
&+&D_k\partial_u \left( u\partial_u u P_k\right)-\lambda_k 
\partial_u\left(u\partial_u P_k\right)\nonumber\\
&-&\lambda_k\partial_u^2\left(uP_k\right)+\alpha_k\partial_u^2 P_k,
\label{fokker}
\ENA
where we have taken into account the fact that, due to homogeneity, 
$\xi$ and $\sigma$ have a zero mean. Here, we dropped the the subscript 
$0$ in $k_0$ and dependence of the all involved quantities on
the scale is simply marked by  the subscript $k$ (the scale dependence
 $\alpha, D$ and $\lambda$ is 
still unspecified).
The stationary solution of (\ref{fokker}) is:
\EQ
P_k(u)=C_k\exp\int_0^u \frac{-\nu_t k^2 y-Dy+\lambda}{Dy^2-2\lambda 
y+\alpha}dy,
\label{statsol}
\EN
where $C_k$ is a normalization constant. The integral appearing in 
(\ref{statsol}) can be explicitly computed in two regimes: in the 
first one,
for $u<<k \nu_0$, we have ($\nu_t=\nu_0$), and we simply get
\EQ
P_{k}(u)=\frac{C}{\left(Du^2-2\lambda u+\alpha\right)^{1/2+\nu_0 k^2/(2D)}},
\nonumber\\
\label{dissipative}
\EN
The range of $u$ for which the PDF follows this algebraic law decreases with
increasing scales. It is the largest
(and hence it is best observed) at the dissipative
scale, where the velocity increments are equivalent to velocity 
derivative or to vorticity. Several remarks are in order about this 
expression. First, notice that the distribution is regularized around 
$u=0$ by the presence of the parameter $\alpha$, but then displays 
algebraic tails. These are well known features of random 
multiplicative process with additive noise (see e.g \cite{nakao98}).
The occurrence of algebraic tails in vorticity PDF's has been noted 
before by \cite{jimenez96,min96} in the context of 2D turbulence. 
However, processes with algebraic tails are characterized by 
divergent moments. These divergences can be removed by taking into 
account finite size effects, like physical upper bounds on the value 
of the process (see, e.g. \cite{laval98} for discussion and 
references) which introduce a cut-off in the probability 
distribution. This effect is automatically taken into account in our 
simple model, via the turbulent viscosity, which prevents unbounded 
growth of velocity fluctuations and introduces an exponential 
cut-off.\

Indeed, in the regime $u>>k\nu_0$, we see that:
\EQ
\frac{d\ln P}{du}\approx \frac{-u\vert u\vert}{D u^2-2\lambda u+\alpha}.
\label{derivat}
\EN
This mean that at large $u$ the PDF decays like an exponential, or 
even faster if $D=0$
(see below). This feature had been noted by Min et al \cite{min96}, 
and finds here a detailed explanation.\

Another important observation is  that the PDF's have an intrinsic 
skewness, which can be traced back to the non-zero value of 
$\lambda$, i.e. to the correlation between the multiplicative and the 
additive noise. The physical picture associated with this correlation 
is related to the correlation between vorticity (present in the large 
scale strain tensor) and stretching (associated with the
term $U\nabla U$, present in $\sigma$),
which is the motor of the energy cascade \cite{andreotti97t}. The 
importance of the additive noise in the skewness generation has been 
stressed before \cite{dubrulle99}. We find here its detailed 
explanation. Note also that the trends towards Gaussian large scale 
behavior of the velocity increments can be easily accounted for if 
the multiplicative noise tends to zero at large scale ($D,\lambda\to 
0$). In such case, the process becomes purely additive, and the 
limiting PDF is Gaussian for $u$ independent turbulent viscosity, or 
can also fall faster than a Gaussian  (like $\exp(-u^3)$,for 
$\nu_t\propto u$). Such a supra Gaussian behavior has been noted 
before
\cite{andreotti97t}.  In between the dissipative and the largest 
scale, the transition operates via PDF's looking like stretched 
exponential.\

This qualitative feature can be tested by comparison with the 
numerical PDF's. Our model, predicts that $d\ln P/du$ should behave 
like the ratio:
\EQ
\frac{d\ln P}{du}=\frac{-u\sqrt{(m_1 u)^2+m_2^2}-m_4 u+m_3}{m_4 
u^2-2m_3 u+m_5}.
\label{fitquad}
\EN
Without loss of generality, we can factorize out the parameter $m_5$. The fit
Therefore only contains four free parameters, which can be easily 
related to the physical parameters of the problem.
  We have computed this derivative for the PDF of longitudinal and 
transverse velocity increments at various scale, and performed the 
four parameters fit.
\begin{figure}[hhh]
\centerline{\psfig{file=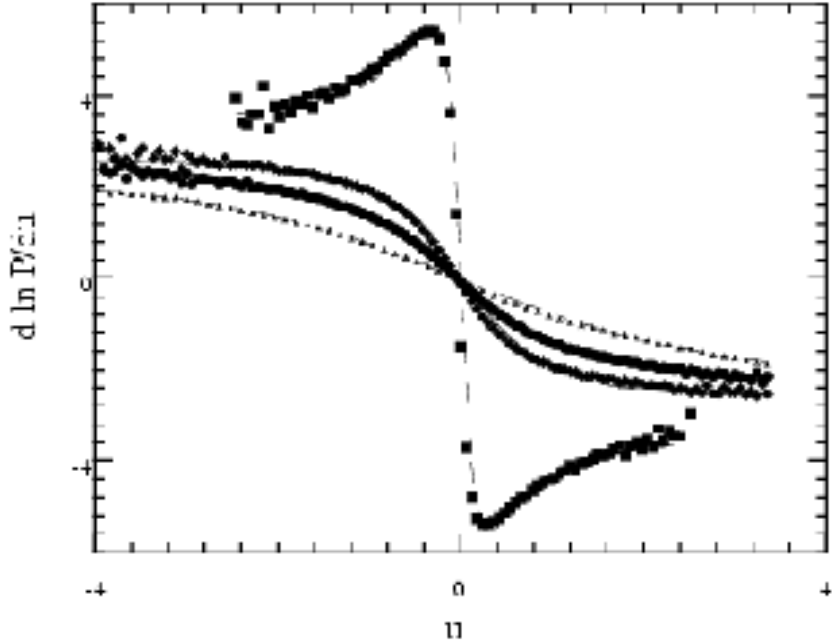,width=12cm}}
\caption[]{Fit of $d\ln P/du$ for the transverse increments, at 
$1/k=2$ (squares),
$1/k=32$ (diamonds), $1/k=42$ (circles) and $1/k=52$ (triangles). The 
fits via formula \ref{fitquad} are given by lines. Note the good 
quality of the fit. Similar fits for other scale separations, and in 
the longitudinal case were obtained.}
\label{fig:deriv-lnp}
\end{figure}
 Examples are shown on Fig. 
\ref{fig:deriv-lnp}. Observe the good quality of the fit, 
but we stress that there is a
rather large uncertainty in the determination of the parameter of the
fits, which sometimes cannot be determined better than up to a factor 
of two by our fit procedure (a standard least square fit). The scale
dependence of the coefficients of the fit, is shown on Fig. 
\ref{fig:coeff-long} and \ref{fig:coeff-trans}.
Note that since transverse velocity increments are symmetrical by 
construction, we have set $m_3=0$ in the fit. Note that in the 
longitudinal case, $m_3$ is smaller than the other coefficients by 
two orders of magnitude. This features the smallness of the skewness, 
and the weakness of the correlation between the multiplicative and 
the longitudinal noise (see Section 2.1). The parameters have a rough 
power law behavior (see Fig. \ref{fig:coeff-long} and 
\ref{fig:coeff-trans}). Theoretically, one expects the ratio 
$m_1/m_2$ to behave like $1/(\nu_0 k)$, if (\ref{viscodimen}) holds. 
The power-law fits of Fig. \ref{fig:coeff-long} and 
\ref{fig:coeff-trans} provides a dependence of $k^{0.53}$ for the 
longitudinal case, and $k^{-0.54}$ for the transverse case, 
corresponding to a scale dependence of $\nu_0$ given by $k^{1.53}$ 
and $k^{0.46}$.

\subsubsection{The Eulerian description}

We now consider (\ref{ray221D}), again in the moving with
the wave packet frame, but now we change the independent variable
from $t$ to $k=k(t)$ which satisfies (\ref{ray221Da}). 
We get
\EQ
\frac{d u}{d\ln k}=-{\cal P}u+\frac{\nu_t k^2}{\xi}-\frac{\sigma_\perp}{\xi},
\label{langevinscale}
\EN
where ${\cal P}$ is a number, accounting for the projection operator 
(a way to mimic higher-dimensional effects in our
1D toy model). Eq. (\ref{langevinscale}) is 
again a Langevin equation for velocity increments in the scale space, 
with multiplicative and additive noises which is now expresses in
an Eulerian form. Similar Langevin processes 
have been proposed before to explain the scale dependence of velocity 
increments \cite{friedrich97,marcq98,marcq00} but without additive 
noise \cite{friedrich98c}.\

\begin{figure}[hhh]
\centerline{\psfig{file=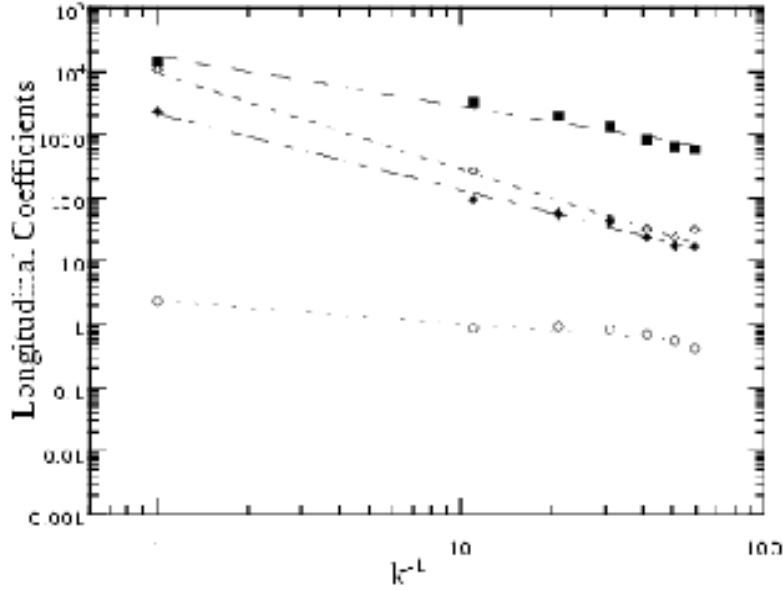,width=12cm}}
\caption[]{Coefficient of the fit of $d\ln(P)/du$ with (\ref{fitquad}) as
a function of $1/k$ for longitudinal velocity increments: $m_1$ 
(squares), $m_2$ (empty diamonds), $m_3$ (circles) and $m_4$ (filled 
diamonds). The lines are the power-law fits: $k^{-0.79}$ (long-dashed 
line); $k^{-1.52}$ (short-dashed line);
$k^{-0.36}$ (dotted line) and $k^{-1.2}$ (dash-dot line). }
\label{fig:coeff-long}
\end{figure}

\begin{figure}[hhh]
\centerline{\psfig{file=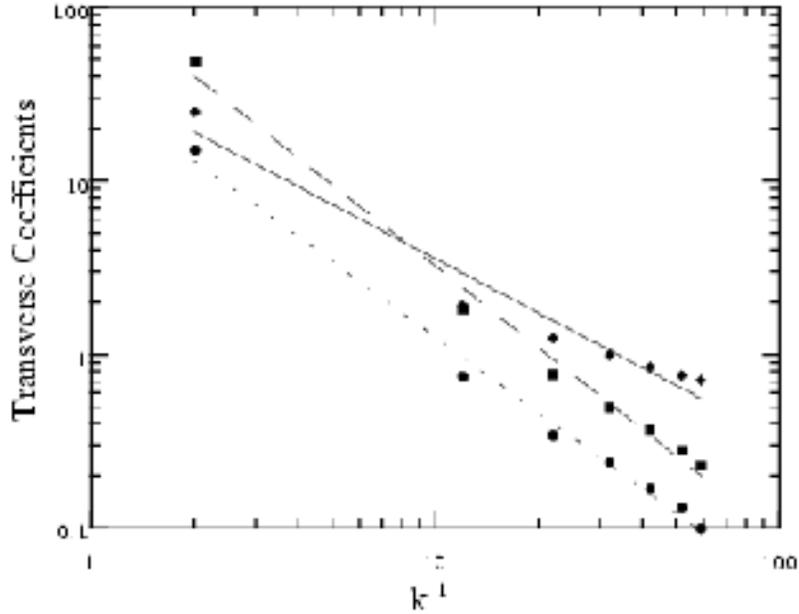,width=12cm}}
\caption[]{Coefficient of the fit of $d\ln(P)/du$ with (\ref{fitquad}) as
a function of $1/k$ for transverse velocity increments: $m_1$ 
(squares), $m_2$ (diamonds), and $m_4$ (circles). The lines are the 
power-law fits: $k^{-1.56}$ (long-dashed line); $k^{-1.04}$ (line);
and $k^{-1.45}$ (dotted line). }
\label{fig:coeff-trans}
\end{figure}

The noises  in this Langevin equation are different from the 
noises appearing in the Lagrangian representation and they would
have a complicated statistics if we assumed that $\sigma$ and $\xi$
were Gaussian in the Lagrangian representation. However, we can
simply assume here that the noises are Gaussian and delta-correlated
in the Eulerian representation (which is different from the
assumption of the previous section) and re-define $\alpha, D$ and
$\lambda$ as
\EQA
<\sigma_{\perp }(k,t(k))\xi^{-1}(k,t(k)) \sigma_{\perp 
}(k',t(k'))\xi^{-1}(k',t(k'))>&=&2\alpha \, \delta(k-k'),\nonumber\\
<\xi^{-1}(k,t(k))\xi^{-1}(k',t(k'))>&=&2D \,  \delta(k-k'),\nonumber\\
<\xi^{-1}(k,t(k))\sigma_{\perp }(k',t(k'))\xi^{-1}(k',t(k'))>&=&2\lambda \, \delta(k-k'),
\label{correforces1}
\ENA
This allows one to derive the Fokker-Planck equation corresponding to 
(\ref{langevinscale}),
\EQA
k\partial_k P(u,k)&=&\partial_u \left({\cal P}u P\right)\nonumber\\
&+&D\partial_u \left( \nu_t k^2 u\partial_u \left[\nu_t k^2 u 
P\right]\right)-\lambda \partial_u\left(\nu_t k^2u\partial_u 
P\right)\nonumber\\
&-&\lambda \partial_u^2\left(\nu_t k^2 uP\right)+\alpha\partial_u^2 P.
\label{fokkerbis}
\ENA
We may use (\ref{fokkerbis}) to derive an equation for the 
moments, by multiplication by $u^n$ and integration over $u$. With 
the shape of the turbulent viscosity given by (\ref{viscodimen}), we 
get:
\EQA
k\partial_k <u^n>&=& - \zeta(n)<u^n>\nonumber\\
&+&nDB^2k^2<u^{n+2}>-\lambda n(2n-1)\nu^2 k^2 
<\frac{ u^{n-1}}{\nu_t}>\nonumber\\
&-&2n^2\lambda B^2 <\frac{u^{n+1}}{\nu_t}>
+\alpha n(n-1)<u^{n-2}>,
\label{momenthiera}
\ENA
where $\zeta(n)=n{\cal P}-n^2Dk^4\nu^2$ is the zero-mode scaling 
exponent. For $n=1$ and taking into account the constraints that 
$<u>=0$ (homogeneity), one get a sort of generalized Karman-Horwath 
equation:
\EQ
<u^3>=\frac{\lambda\nu^2}{DB^2} 
<\frac{1}{\nu_t}>+\frac{2 \lambda}{Dk^2}<\frac{u^2}{\nu_t}>.
\label{Khreplace}
\EN
As in the Lagrangian case, this means that skewness (related to 
non-zero $<u^3>$ is generated through non-zero value of $\lambda$, 
i.e. through correlations of the multiplicative and the additive 
noise.
However, due to the turbulent viscosity, we cannot explicitly solve 
the hierarchy of equation. In many homogeneous turbulent flows, 
however, the skewness
(proportional to $\lambda$) is quite small, and moment of order $2n+1$ are
generally negligible in front of moments of order $2n+2$.
For {\sl even} moments, this remark suggest that to first order in 
the skewness, and for $2n>1$ the dynamics is simply given by:
\EQA
k\partial_k <u^{2n}>&=&-\zeta(2n)<u^{2n}>
+2nDB^2k^2<u^{2n+2}>\nonumber\\
&+&\alpha 2n(2n-1)<u^{2n-2}>+O(\lambda^2).
\label{approxi}
\ENA
Note that
$D/\alpha$ is given by the parameter $m_4$ in our fit, and is such that
$Dk^2/\alpha$ increases with $k$ (see Figs. \ref{fig:coeff-long} and \ref{fig:coeff-trans}).
 Therefore, at small
scales, the dominant balance is:
\EQ
k\partial_k <u^{2n}>=
2nDB^2k^2<u^{2n+2}>.
\label{small}
\EN
The solution is $<u^{2n}>\propto k^{-2n}$. This is the usual "regular" scaling
in the dissipative zone.
For larger scales,
$DB^2k^2\ll \alpha$, and if $\alpha$ varies like a power law, the 
general solution of (\ref{approxi}) is a sum of power-laws:
\EQ
<u^{2n}>=\sum_{p=0}^n a_p \alpha^{n-p} k^{-\zeta(2p)}.
\label{genesol}
\EN
This solution illustrates the famous mechanism of "zero-mode 
intermittency" \cite{gawedski95}. Here, the zero mode is the solution 
of the homogeneous part of  eq. (\ref{approxi}), i.e. a power-law of 
exponent $-\zeta(2n)$. Without the zeroth mode, responsible for the 
first
$n-1$ scaling laws, the $2n$-th moment will scale in general like 
$\alpha(k)^n$ (provided one assume that this dominates the other 
term), i.e. will be related to the turbulent forcing. This is the 
standard Kolmogorov picture. When the zeroth mode is taken into 
account, the moment now includes new power laws,
whose exponent is independent of the external forcing, and which can 
be dominant in the inertial range, thereby causing anomalous scaling. 
In the present case, the scaling exponent is quadratic in $n$, and 
reflects the log-normal statistics induced by the Gaussian 
multiplicative noise, in agreement with the latest wavelet analysis 
of Arneodo et al \cite{arneodo99}. Note also that the competition 
between the zeroth mode scaling and the scaling due to external 
forcing forbids the moments to scale like a power-law, thereby 
generating a breaking
of the scale symmetry.\

\begin{figure}[hhh]
\centerline{\psfig{file=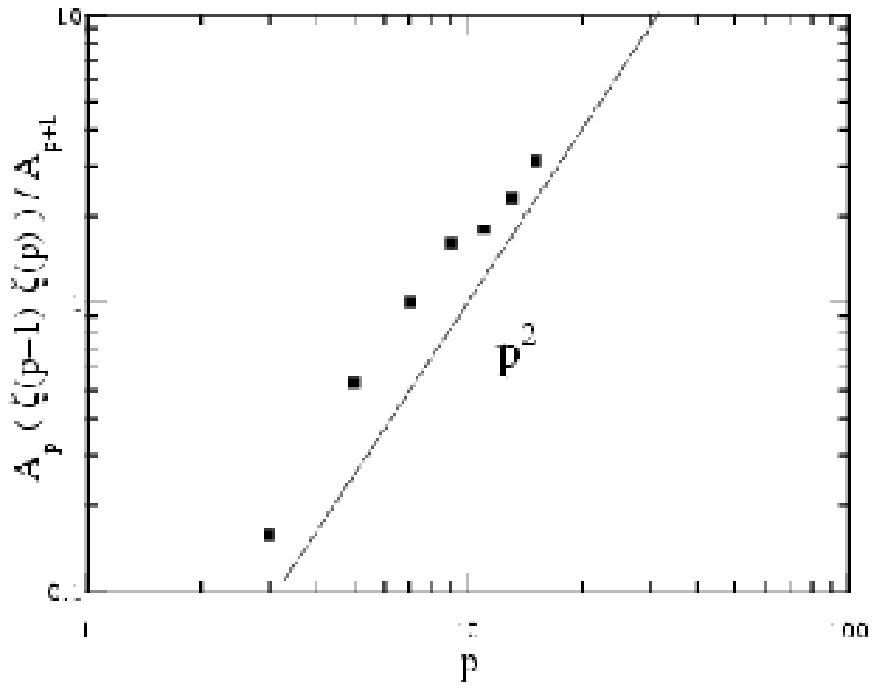,width=12cm}}
\caption[]{$(\zeta(2n+2)-\zeta(2n+1))A_{2n+1}/A_{2n+2}$ as a function of
$n$ in a high Reynolds number boundary layer. Here $A_{n}$ is the 
prefactor of the (non-dimensional) structure function of order $n$. 
The line is $n^2$, the
prediction of our toy model. }
\label{fig:prefactor}
\end{figure}

For {\sl odd} moments, we cannot perform any rigorous expansion 
because all the terms of the equation are of order $\lambda$. For low 
order moments, however, the computation of $<u^{n+1}/\nu_t>$ mainly 
involve velocity increments close to the center of the distribution, 
for which $\nu_t\approx \nu_0$. So, for low
order, it is tempting to approximate the equation for the odd-order moments
 by
\EQ
k\partial_k <u^{2n+1}>\approx -\zeta(2n+1)<u^{2n+1}>-2n^2 \lambda B^2 
<u^{2n+2}>\nu^{-1}.
\label{approloose}
\EN
This approximation is only valid in the inertial range, where the 
last term of (\ref{momenthiera}) can be neglected. An immediate 
consequence of this loose approximation is that 
$\zeta(2n+1)=\zeta(2n+2)$ in the inertial range. Odd moments (without 
absolute values)
are very difficult to measure because of cancelation effects which 
introduce a lot of
noise. In  our case, due to our limited inertial range, we cannot 
compute these exponent with a sufficient degree of accuracy.
\begin{figure}[hhh]
\centerline{\psfig{file=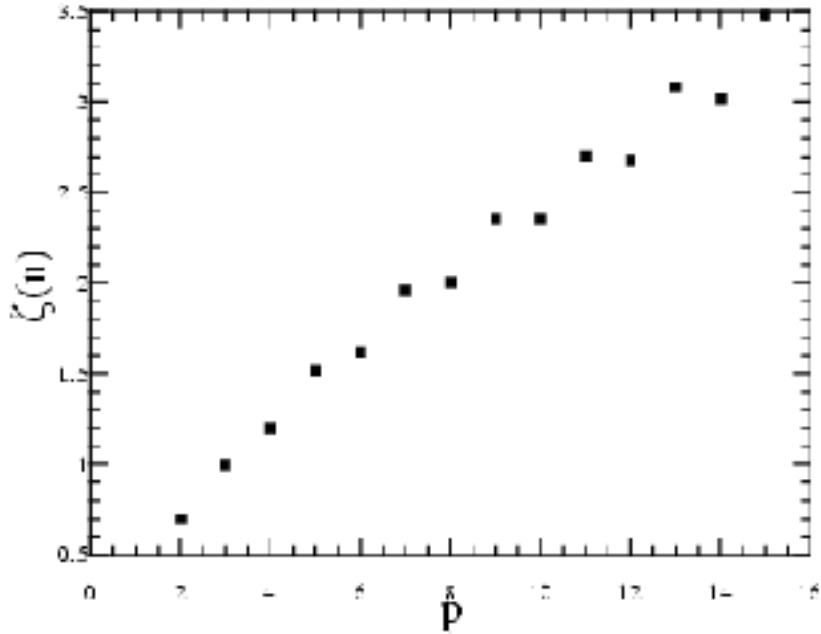,width=12cm}}
\caption[]{Exponents of the structure function in a high Reynolds 
number boundary layer \cite{dhruva97}. Note the tendency for 
$\zeta(2n+1)=\zeta(2n+2)$
for $n>3$. }
\label{fig:exposants}
\end{figure}
 A 
careful investigation performed in a high Reynolds number boundary 
layer \cite{stolovitsky93} however seem to be in agreement with our 
prediction, as is shown in Fig. \ref{fig:exposants}: it is striking to 
observe that $\zeta(5)\approx \zeta(6)$, $\zeta(7)\approx \zeta(8)$, 
etc making the curve look as if odd and even scaling exponents are 
organized on a separate curve \cite{zubair93}. A second independent
experimental check of our prediction (\ref{approloose}) is that
$(\zeta(2n+2)-\zeta(2n+1))<u^{2n+1}>/<u^{2n+2}>$ should scale, for 
$n>3$ like $n^2$. Fig. \ref{fig:prefactor}
shows that this is indeed the case.\
\section{Discussion}
In this paper, we have shown that non-local interaction are
responsible for intermittency corrections in the statistical
behavior of 3D turbulence. Removal of the local interaction
in numerical simulations leads to a substantial increase in
the number of the tornado-like intense vortex filaments and
to a stronger anomalous corrections in the higher cumulants
of the velocity increments. It is also accompanied by
  a modification of
the energy transfer in the inertial range, tending to create a
flatter energy spectrum.
The intermittency corrections and the spectra are close to that observed
in high Reynolds number boundary layer, suggesting that
the non-local interactions prevail in this geometry. This could be explained
by presence of the mean flow, which geometrically favors non-local triads in
the Fourier transform of the non-linear interactions.
We showed that replacing the removed local interactions by a simple
turbulent viscosity term allows to restore the correct intermittency
and the energy characteristics. Our results agree with the belief
that intermittency is related to thin  vortices
amplified by the external large scale strain similarly to the classical
Burgers vortex solution. Local interactions can be viewed at mutual
interaction of these thin intense vortices which result in their
destruction, which is also in agreement with our results.

To prove that the enhanced intermittency is not simply the result
of the stronger small-scale observed in the RDT simulation,
we performed yet another numerical experiment in which the non-local
interactions were neglected and only the local ones retained.
This resulted in even higher (than in RDT) level of the small scales
but it exhibited much less intermittency, which confirms our view
that non-locality is crucial for generation of the intermittent
structures.

The result that the net effect of the local interactions is
to destroy the intermittent structures is at odds with a very common
belief that the intermittency is due to the vortex reconnection
process which takes a form of a finite time vorticity blow-up.
Indeed, the later is a strongly nonlinear process in which the local
vortex-vortex interactions are important. However, this process seems
to be dominated by another local processes the net result of which
is to destroy the high-vorticity structures rather than to create them.
It would be premature to claim, however, that the same is true at
any arbitrarily high Reynolds number.

Our numerical approach sets severe limitations to the value of
the Reynolds number we are able to explore. In this context, it is interesting
  to point out that preliminary tests regarding the importance of
non-local interactions
  have been conducted on a velocity field coming from a
very large Reynolds number boundary layer \cite{carlier00}. Even though
the test is not complete (the probes only permit the accurate
measurement of special components of the velocity field), it tend
to suggest that non-local interactions dominate the local interactions
by several orders of magnitude.
Our results would also explain the findings of the Lyon
team \cite{pinton98},
who found that when probing fluid area closer and closer to
a large external vortices,
or to a wall boundary, one could measure energy spectra moving from a
$k^{-5/3}$ law towards a $k^{-1}$ spectra, while anomalous corrections
in scaling exponents would become more pronounced. At the light of our study,
this could be simply explain by a trend towards more non-local dynamics via
the mean-shear effects at the wall.

Based on the conclusions of our numerical study, we developed a new 
toy model of turbulence to study the intermittency. It has the form
of a Langevin equation for the velocity increments with coupled 
multiplicative and additive noise. We showed how this toy model could 
be used to understand qualitatively certain observed features of 
intermittency and anomalous scaling laws. Among other things, we 
showed how the coupling between the two forces is related to the 
skewness of the distribution, and how algebraic and stretched 
exponential naturally arise from the competition between the 
multiplicative and the additive noise. We tested our qualitative 
predictions with experimental and numerical data, and found good 
general agreement.  To be able to turn our toy model into a tool for 
"quantitative" study of the intermittency, several developments are 
needed. The first one is to consider the multi-dimensional version 
of our toy model, to be able to couple longitudinal and transverse 
increments. The scale dependence of the turbulent viscosity and of 
the forcing needs to be further investigated, possibly using tools 
borrowed from the Renormalization Group theory (see e.g. 
\cite{canuto96}). Also, the non-Gaussianity of the noises could be 
taken into account.\

In 1994, Kraichnan \cite{kraichnan94} proposed an analytically 
solvable, new toy model for the passive scalar, which provided a 
substantial increase of our understanding of the passive scalar 
intermittency. Our model, built using the non-local hypothesis, is a 
direct heir of this philosophy in that, as in passive scalars,
 all important intermittency
effects are produced via a linear dynamics. The nonlinear (local)
scale interactions are important too because they are the main carrier
of the energy cascade, but it is only their mean effect and not
statistical details that are essential.

the present paper.

\vspace{2cm}

{\bf Acknowledgments}\

BD acknowledges the support of a NATO fellowship and 
J-P Laval is thankful for the support of the 
{\it Laboratoire de M\'ecanique de Lille}, France.
SN acknowledges the support of the TMR European network grant
``Intermittency in Turbulent Systems'' (ERB FMR XCT 98-0175).
SN thanks Bob Kerr for useful discussion of the turbulent
viscosity models. We thank Keith Moffatt, Oleg Zaboronki and our
referees for suggesting an additional ``local'' numerical test
which significantly reinforced our results. 

\bibliographystyle{plain}
\bibliography{../../Biblio/biblio}

\end{document}